\documentclass[12pt]{article}
\usepackage{a4wide}
\usepackage{amssymb}
\usepackage{graphicx}
\usepackage{epsfig}
\usepackage[tbtags]{amsmath}
\usepackage{natbib}

\begin{document}
{\renewcommand{\thefootnote}{\fnsymbol{footnote}}
\hfill  IGC--11/1--2\\
\medskip
\begin{center}
{\LARGE  Loop Quantum Gravity and Cosmology:\\[3mm] A dynamical introduction}\\
\vspace{1.5em}
Martin Bojowald\footnote{e-mail address: {\tt bojowald@gravity.psu.edu}}
\\
\vspace{0.5em}
Institute for Gravitation and the Cosmos,\\
The Pennsylvania State
University,\\
104 Davey Lab, University Park, PA 16802, USA\\
\vspace{1.5em}
\end{center}
}

\setcounter{footnote}{0}

\begin{abstract}
 Loop quantum gravity and cosmology are reviewed with an emphasis on
 {\em evaluating} the dynamics, rather than constructing it. The three
 crucial parts of such an analysis are (i) deriving effective
 equations, (ii) controlling the theory's microscopic degrees of
 freedom that lead to its spatial discreteness and refinement, and (iii)
 ensuring consistency and anomaly-freedom. All three issues are
 crucial for making the theory testable by conceptual and
 observational means, and they remain challenging. Throughout this
 review, the Hamiltonian nature of the theory will play a large role
 for properties of space-time structure within the framework
 discussed.
\end{abstract}

\vspace{1cm}

\begin{quote}
  ``{\em It would be permissible to look upon the Hamiltonian form as the
  fundamental one, and there would then be no fundamental
  four-dimensional symmetry in the theory. One would have a
  Hamiltonian built up from four weekly [sic] vanishing functions,
  given by [the Hamiltonian and diffeomorphism constraints]. The usual
  requirement of four-dimensional symmetry in physical laws would then
  get replaced by the requirement that the functions have weakly
  vanishing P.b.'s, so that they can be provided with arbitrary
  coefficients in the equations of motion, corresponding to an
  arbitrary motion of the surface on which the state is defined.}''

  \flushright {\sc P.A.M.~\cite{DiracHamGR}}
\end{quote}

\vspace{1cm}

\mbox{}

\section{Introduction}

In its different incarnations, quantum gravity must face a diverse set
of fascinating problems and difficulties, a set of issues best seen as
both challenges and opportunities. One of the main problems in
canonical approaches, for instance, is the issue of anomalies in the
gauge algebra underlying space-time covariance. Classically, the gauge
generators, given by constraints, have weakly vanishing Poisson
brackets with one another: they vanish when the constraints are
satisfied. After quantization, the same behavior must be realized for
commutators of quantum constraints (or for Poisson brackets of
effective constraints), or else the theory becomes inconsistent due to
gauge anomalies. If and how canonical quantum gravity can be obtained
in an anomaly-free way is an important question, not yet convincingly
addressed in full generality. Posing one of the main obstacles to a
complete formulation of quantum gravity, this issue is hindering
progress toward a detailed evaluation of quantum gravitational
dynamics. A reliable phenomenological analysis must, after all, start
with a consistent set of sufficiently general dynamical equations.

But the strong and tough requirement of anomaly-freedom is also an
opportunity, for it allows an analysis of quantum space-time and the
changes in its structure possibly implied by quantum gravity.
Addressing the anomaly problem is, moreover, crucial for an
understanding of the dynamics of quantum gravity, both in the sense of
{\em constructing} consistent dynamical equations at the quantum level
and in the sense of {\em evaluating} equations and their solutions to
bring out physical effects.

Although the anomaly problem has not been addressed in full
generality, several model systems have by now been analyzed in loop
quantum cosmology, as reviewed by \cite{LivRev}, with this question in
mind. Loop quantum cosmology is a rather wide area within loop quantum
gravity, analyzing several classes of model systems and perturbations
around them. Loop quantum gravity, detailed by \cite{Rov},
\cite{ALRev} and \cite{ThomasRev}, is a canonical quantization of
general relativity based on holonomies (the eponymous loops) as
elementary variables. The use of holonomies allows a
background-independent formulation free of auxiliary metrics, and it
implies several specific properties of the resulting dynamics.

In all the systems used in loop quantum cosmology, quantization
techniques close to those of a general loop quantization are used;
they can thus be seen as capturing at least some of the crucial
properties of full loop quantum gravity. To different degrees, most of
these models make additional use of symmetry reduction as introduced
by \cite{SymmRed}, simplifying much of the quantum geometry and
thereby providing rather direct access to the much less understood
quantum dynamics.  Thanks to these steps, implications of the quantum
dynamics of loop quantum gravity, as generally defined based on
\cite{RS:Ham} and \cite{QSDI}, have been evaluated quite explicitly
for the first time.

In general terms as well as for particular questions arising in loop
quantum cosmology and loop quantum gravity, three key issues regarding
quantum space-time arise, not unlike what one would expect for any
fundamental gauge theory with microscopic degrees of freedom:
\begin{description}
\item[Effective dynamics:] In quantum gravity,
  geometry is described unsharply by whole states with all their
  fluctuations and correlations, in addition to the expectation values
  for an average geometry. Equations of motion for expectation values
  receive quantum corrections in their effective dynamics, as it may
  describe a quantum geometry. Along with this consequence of
  quantizing gravity come not only new mathematical space-time
  structures but also a vast enlargement of the number of degrees of
  freedom by quantum variables.

  The strongest control of such a high-dimensional dynamical quantum
  system is usually obtained for dynamical coherent states, defined as
  states saturating uncertainty relations at all times. If such states
  exist, they provide insights into the minimal deviations from
  classical behavior expected for a quantum system.  The form and
  behavior of dynamical coherent states in loop quantum cosmology can
  be highlighted in several models, bringing out the role of
  space-time fluctuations and correlations as degrees of freedom
  beyond the classical ones. Exact dynamical coherent states exist
  only in special models and for specific initial values.
  Nevertheless, they allow interesting views on the generic quantum
  dynamics as it arises in quantum gravity.

  Before all corrections are derived for a large class of models, a
  clear analysis provided by dynamical coherent states, when they
  exist, unambiguously shows the first deviations from classical
  behavior. More generally, when exact dynamical coherent states do
  not exist, additional quantum corrections will result. They can
  often be computed perturbatively, analogously to loop corrections in
  interacting quantum field theories.
\item[Discrete dynamics:] In addition to those generic effects due to
  non-classical state parameters, underlying space-time structures
  typically change even for the expectation values of a
  quantum-gravity state. Most importantly, discrete geometry, at least
  in purely spatial terms which is by now well understood in loop
  quantum gravity following \cite{AreaVol,Area,Vol2}, shows several
  detailed properties of importance for the dynamics and thus for
  space-time geometry.

  A spatial slice in space-time is equipped with a discrete quantum
  geometry, roughly seen as making space built from atomic patches of
  certain discrete sizes. One of the main problems of quantum dynamics
  is to show how these spatial atoms along slices fit together to form
  a quantum space-time --- or, more dynamically, how the spatial atoms
  change, merge, subdivide and interact as time is let loose.
  For an expanding universe, one would
  expect the discrete spatial structure to be refined as the volume
  increases; otherwise discrete sizes would be enlarged by huge
  factors, especially during inflation, making them macroscopic.  The
  full dynamics of loop quantum gravity has indeed provided several
  hints that the number of discrete building blocks must change from
  slice to slice, once the dynamics is implemented consistently. This
  lattice refinement can be modelled even in the simplest, most highly
  symmetric situations of loop quantum cosmology, laid out by
  \cite{InhomLattice,CosConst}. And it has shown several specific
  implications by which its precise form can already be constrained.
\item[Consistent dynamics:] Dynamics unfolds in
  time, but time is relative. Making sure that descriptions using
  different notions of time, corresponding to measurements by
  different observers, can agree about their physical insights
  requires the consistency conditions of general covariance. Since
  general covariance in gravity is implemented by gauge
  transformations, any quantization or even just a modification of the
  theory must, for consistency, respect this principle and be anomaly
  free.

  While the previous two points manifest themselves already in
  homogeneous models, where they can most easily be studied, the
  anomaly problem arises only in inhomogeneous situations. Within
  homogeneity, all but one of the gauge transformations underlying
  covariance are fixed. Anomalies can only arise if there are at least
  two independent gauge transformations; after quantization, they
  would be anomalous if their composition is no longer a gauge
  transformation.  One could, of course, make it a gauge
  transformation by definition, by declaring the whole group generated
  by the independent gauge transformations as the gauge group. But if
  this group is too large, it would identify variables which are to be
  considered physically distinct, removing observables and degrees of
  freedom and in many cases leaving no non-trivial solutions. Changing
  the gauge transformations of a classical theory by quantum effects
  requires much care; only so-called consistent deformations of the
  classical gauge generators can provide well-defined quantizations.
\end{description}

Addressing these questions is crucial, not only for a complete
formulation of quantum gravity but also for reliable cosmological
applications based on the resulting set of equations (such as
singularity removal or structure formation). In what follows, we will
describe the current status based on the models of loop quantum
cosmology.

\section{Effective dynamics}

In a general sense, effective equations of a quantum system describe
the behavior of expectation values in a state. Deriving such equations
for the expectation values of basic operators, such as $\hat{q}$ and
$\hat{p}$ in quantum mechanics, shows how quantum effects change the
classical equations of motion. If the equations can be solved or
analyzed, the manifestation of quantum behavior will be seen. Such
effects play an important role in any interacting quantum theory, so
also in quantum gravity and especially in quantum cosmology which
devotes itself to the analysis of extremely long evolution
times. During those times, quantum states may change drastically, and
quantum corrections grow to be significant.

\subsection{Momentous quantum mechanics}

Effective equations are state dependent since they are equations for
expectation values in a state with many more independent
parameters. In general, quantum fluctuations will influence the
behavior of expectation values, and must then be included in effective
equations in some form. More generally, and following \cite{EffAc}, we
can parameterize a state by all its moments
\begin{equation}\label{Moments}
G^{a,b}\equiv G^{\underbrace{\scriptstyle q\cdots q}_a\underbrace{\scriptstyle p\cdots p}_b}=\langle(\hat{q}-\langle\hat{q}\rangle)^a
 (\hat{p}-\langle\hat{p}\rangle)^b\rangle_{\rm Weyl}
\end{equation}
defined for $a+b\geq 2$, in addition to the expectation values
$\langle\hat{q}\rangle$ and $\langle\hat{p}\rangle$. (The subscript
``Weyl'' denotes totally symmetric ordering. We will use the two
notations indicated on the left interchangeably, at least for small
$a+b$.) For instance, $G^{2,0}\equiv G^{qq}$ is the square of position
fluctuations, and $G^{1,1}\equiv G^{qp}=
\frac{1}{2}\langle\hat{q}\hat{p}+\hat{p}\hat{q}\rangle-
\langle\hat{q}\rangle \langle\hat{p}\rangle$ the covariance.  The
values of moments are not completely arbitrary, most importantly being
restricted by uncertainty relations such as
\begin{equation} \label{Uncert}
 G^{qq}G^{pp}- (G^{qp})^2\geq \frac{\hbar^2}{4}\,.
\end{equation}
For pure states, the set of moments as defined here is overcomplete;
the framework more generally allows for mixed states, too.

All the moments are dynamical. Given a Hamiltonian operator, for every
observable $\hat{O}$ we have an equation of motion 
\begin{equation} \label{dOdt}
 \frac{{\rm
  d}\langle\hat{O}\rangle}{{\rm d}t}=
\frac{\langle[\hat{O},\hat{H}]\rangle}{i\hbar}
\end{equation}
of its expectation value.  Specific examples are obtained for the
terms in a moment, and so we can derive their equations of motion. For
the square of position fluctuations, for instance, we have
\[
 \frac{{\rm d} G^{qq}}{{\rm d}t}= \frac{{\rm d}}{{\rm d}t}
     (\langle\hat{q}^2\rangle - \langle\hat{q}\rangle^2)=
     \frac{\langle[\hat{q}^2,\hat{H}]\rangle}{i\hbar}-
     2\langle\hat{q}\rangle
     \frac{\langle[\hat{q},\hat{H}]\rangle}{i\hbar}\,.
\]
Introducing Poisson brackets on the space of moments via
\begin{equation} \label{AB}
 \{\langle\hat{A}\rangle,\langle\hat{B}\rangle\}=
\frac{\langle[\hat{A},\hat{B}]\rangle}{i\hbar}\,,
\end{equation}
extended by linearity and the Leibniz rule, equations of motion take
Hamiltonian form:
\begin{equation} \label{Gdot}
 \frac{{\rm d}G^{a,b}}{{\rm d}t}= \{G^{a,b},H_Q\}
\end{equation}
with the quantum Hamiltonian $H_Q:= \langle \hat{H}\rangle$.

In general, the moments are all coupled to one another and to the
equations for expectation values. One can see this by expanding
\begin{eqnarray}
 H_Q(\langle\hat{q}\rangle,\langle\hat{p}\rangle, G^{a,b})&=&
\langle H(\hat{q},\hat{p})\rangle=\langle
H(\langle\hat{q}\rangle+(\hat{q}-\langle\hat{q}\rangle),
\langle\hat{p}\rangle+(\hat{p}-\langle\hat{p}\rangle))\rangle \nonumber\\
&=& H(\langle\hat{q}\rangle,\langle\hat{p}\rangle)+\sum_{a,b:a+b\geq 2} 
\frac{1}{a!b!}
\frac{\partial^{a+b}H(\langle\hat{q}\rangle,\langle\hat{p}\rangle)}{\partial \langle\hat{q}\rangle^a\partial \langle\hat{p}\rangle^b}G^{a,b}
\end{eqnarray}
(where we assumed the Hamiltonian operator $\hat{H}$ to be
Weyl-ordered in $\hat{q}$ and $\hat{p}$) and noticing the coupling
terms of expectation values and moments for any non-quadratic
potential. The Hamiltonian flow (\ref{dOdt}) or (\ref{Gdot}) then
couples expectation values $\langle\hat{q}\rangle$ and
$\langle\hat{p}\rangle$ to all the moments.

At this stage, we have an exact but usually horribly complicated
Hamiltonian description of quantum evolution. The partial differential
equation for a state in Schr\"odinger's formulation is replaced by
infinitely many ordinary differential equations for the moments. Most
of the labor that goes into deriving tractable effective equations
consists in extracting the required information about expectation
values without having to solve for a full state, or all its moments,
and to specify the regimes where this is reliable.  In a semiclassical
approximation based on near-Gaussian states, for instance, a moment of
order $a+b$ is typically of the order $\hbar^{(a+b)/2}$, giving rise
to a natural expansion in powers of $\hbar$. To any given order, only
finitely many moments need be considered.\footnote{Truncating the
  phase space in this way leads to Poisson manifolds spanned by the
  moments to a certain order. In general, these Poisson structures are
  degenerate; for instance, there are three independent second order
  moments, forming a space which cannot carry a non-degenerate Poisson
  structure.  Effective equations thus make crucial use of Poisson
  geometry, not symplectic geometry as in other geometric formulations
  of quantum mechanics such as the one going back to
  \cite{GeomQuantMech}.}

\subsection{Harmonic oscillator}

In special systems, equations of motion for the moments decouple
automatically to finite sets. The best known case is the Harmonic
oscillator, whose quantum Hamiltonian is
\begin{equation} \label{HQHarm}
H_Q=\frac{1}{2m}\langle\hat{p}\rangle^2+
\frac{1}{2}m\omega^2\langle\hat{q}\rangle^2
+\frac{1}{2}m\omega^2G^{qq}+\frac{1}{2m}G^{pp}\,.
\end{equation}
Second-order moments appear, but they do not couple to the expectation
values. They rather provide the zero-point energy due to quantum
fluctuations. Hamiltonian equations of motion are only finitely
coupled,
\begin{eqnarray}
\nonumber \frac{{\rm d}\langle{\hat{q}}\rangle}{{\rm d}t}&=&
\{\langle\hat{q}\rangle,H_Q\}=\frac{1}{m} \langle\hat{p}\rangle\\
\frac{{\rm d}\langle{\hat{p}}\rangle}{{\rm d}t}&=&
\{\langle\hat{p}\rangle,H_Q\}=-m\omega^2 \langle\hat{q}\rangle
\\\nonumber
\frac{{\rm d} G^{a,b}}{{\rm d}t}&=&\{{G}^{a,b},H_Q\}=
\frac{1}{m}aG^{a-1,b+1}-m\omega^2 bG^{a+1,b-1}\,.
\end{eqnarray}
For the expectation values we have exactly the classical equations
without quantum corrections, as is well-known for the harmonic
oscillator. Here, effective equations for $\langle\hat{q}\rangle$ and
$\langle\hat{p}\rangle$ are identical to the classical ones.

The remaining equations then show how the state evolves once initial
values for moments have been chosen. For stationary states, for
instance, a vanishing covariance $G^{qp}$ ensures that fluctuations
are time-independent. The covariance is constant in time if
$\dot{G}^{qp}= G^{pp}/m- m\omega^2G^{qq}=0$, or $G^{pp}=
m^2\omega^2G^{qq}$. With this, the non-classical contribution to the
energy $H_Q$ in (\ref{HQHarm}) is $m\omega^2G^{qq}$, where $G^{qq}\geq
\hbar/2m\omega$ from (\ref{Uncert}). A minimal two-point energy
$\frac{1}{2}\hbar\omega$ results for the ground state, derived purely
by effective means (even though the ground state is not at all
semiclassical).

Decoupled equations of this form are very useful because they can
directly show important aspects of quantum dynamics. Expectation-value
equations decouple from the rest whenever the algebra of basic
operators together with the Hamiltonian is linear: in this case, the
time derivative of the expectation value of any one of the basic
operators is an expectation value of a basic operator. Such systems
are solvable in a strong sense; there is no quantum back-reaction from
the moments on the dynamics of expectation values. And the dynamics of
moments of a given order depends only on moments of the same order. In
terms of quantum field theory, solvable models correspond to free
theories.

\subsection{Low-energy effective potential}

With interactions, or for anharmonic terms, quantum back-reaction
results. Moments couple non-trivially to expectation values and become
important for their dynamics. In this way, the state dependence of
effective equations ensues. The system of equations necessarily
becomes of higher dimension than classically, with new dynamical
quantum degrees of freedom given by the moments. 

For a quantum mechanical system with an anharmonic potential, for
instance, we have the classical Hamiltonian $H=
\frac{1}{2m}p^2+\frac{1}{2}m\omega^2q^2+U(q)$ with anharmonicity
$U(q)$. In terms of dimensionless quantum variables
\begin{equation} \label{Gtilde}
 g^{a,b}=\hbar^{-(a+b)/2}(m\omega)^{a/2-b/2}G^{a,b}
\end{equation}
the quantum Hamiltonian is
\begin{equation} \label{HQexpand}
H_Q=\frac{1}{2m}\langle\hat{p}\rangle^2+\frac{1}{2}m\omega^2
\langle\hat{q}\rangle^2+U(\langle\hat{q}\rangle)
+\frac{\hbar\omega}{2}(g^{2,0}+g^{0,2})+\sum_{n=2}^{\infty}\frac{1}{n!}
\left(\frac{\hbar}{m\omega}\right)^{n/2}
U^{(n)}(\langle\hat{q}\rangle)g^{n,0}
\end{equation}
and generates equations of motion
\begin{eqnarray}
\nonumber  \frac{{\rm d}\langle\hat{q}\rangle}{{\rm d}t}&=& \frac{1}{m}\langle\hat{p}\rangle\label{eom}\\
\frac{{\rm d}\langle\hat{p}\rangle}{{\rm
    d}t}&=&-m\omega^2\langle\hat{q}\rangle
-U'(\langle\hat{q}\rangle)-\sum_{n=2}^{\infty}\frac{1}{n!}
\left(\frac{\hbar}{m\omega}\right)^{n/2}U^{(n+1)}(\langle\hat{q}\rangle)
g^{n,0}\\
\nonumber \dot{g}^{a,b}&=&a\omega
g^{a-1,b+1}-b\omega g^{a+1,b-1}
-\frac{bU''(\langle\hat{q}\rangle)}{m\omega}g^{a+1,b-1}\\
\nonumber&&+
\frac{b\sqrt{\hbar}U'''(\langle\hat{q}\rangle)}{2(m\omega)^{3/2}} 
\left(g^{a,b-1}g^{qq}- g^{a+2,b-1}+ \frac{(b-1)(b-2)}{12}g^{a,b-3}\right)\\
\nonumber&&
+\frac{b\hbar U''''(\langle\hat{q}\rangle)}{6(m\omega)^2}
\left(g^{a,b-1}g^{qqq}- g^{a+3,b-1}+ \frac{(b-1)(b-2)}{4} g^{a+1,b-3}\right)
+\cdots\,.
\end{eqnarray}

The leading quantum correction appears in the equation for
$\langle\hat{p}\rangle$ at order $\hbar$ ($n=2$), for which we have to
know the position fluctuation $g^{2,0}=g^{qq}$. In general, this is an
independent variable subject to its own equation of motion. Its
evolution couples to other quantum variables, eventually making the
whole infinite system coupled. At this stage, approximations are
required. For semiclassical states, we may drop terms of higher order
in $\hbar$,\footnote{By definition (\ref{Gtilde}), the leading
semiclassical $\hbar$-dependence of $g^{a,b}$ has been factored out,
such that explicit factors of $\hbar$ in the equations of motion
suffice to read off orders. In terms of the original moments, at
second order we ignore all terms $\hbar^nG^{a,b}$ with $2n+a+b>2$.}
providing a closed system of effective equations
\begin{eqnarray*}
  \frac{{\rm d}\langle\hat{q}\rangle}{{\rm d}t}&=& 
\frac{1}{m}\langle\hat{p}\rangle\\
\frac{{\rm d}\langle\hat{p}\rangle}{{\rm d}t}&=&-m\omega^2\langle\hat{q}\rangle -U'(\langle\hat{q}\rangle)-\frac{1}{2}\frac{\hbar}{m\omega}
U'''(\langle\hat{q}\rangle)g^{qq}+ O(\hbar^{3/2})\\
\frac{{\rm d}g^{qq}}{{\rm d}t}&=& 2\omega g^{qp}\\
\frac{{\rm d}g^{qp}}{{\rm d}t}&=&
\omega(g^{pp}-g^{qq})- \frac{U''(\langle\hat{q}\rangle)}{m\omega}
g^{qq}+ O(\sqrt{\hbar})\\
\frac{{\rm d} g^{pp}}{{\rm d}t}&=& -2\omega g^{qp}- 2
\frac{U''(\langle\hat{q}\rangle)}{m\omega} g^{qp}+O(\sqrt{\hbar})\\
\end{eqnarray*}
for expectation values and second-order moments. With higher moments
dropped, this system shows the leading quantum corrections in a
finitely-coupled system.

For anharmonic oscillators, a further treatment is possible which
makes use of an adiabatic approximation: we assume that time
derivatives of the moments are small compared to the other terms. In
this case, equations of motion for moments become algebraic
relationships between them. For the leading adiabatic order
$g_0^{a,b}$, combined with the previous $\hbar$-expansion, we must
solve
\[
 0=\{g^{a,b}_0,H_Q\}=\omega\left(ag_0^{a-1,b+1}-b
 \left(1+\frac{U''(\langle\hat{q}\rangle)}{m\omega^2}\right)
g_0^{a+1,b-1}\right)+O(\sqrt{\hbar})
\]
which is of the stationary harmonic form, but with position-dependent
coefficients. Although ${\rm d}\langle\hat{q}\rangle/{\rm d}t\not=0$,
this is consistent with the adiabatic assumption. The general solution
is
\[
 g_0^{a,b}= { (a+b)/2 \choose b/2} {a+b \choose b}^{-1}\left(
 1+\frac{U''(\langle\hat{q}\rangle)}{m\omega^2}\right)^{b/2}g^{a+b,0}_0
\]
for even $a$ and $b$, and $g_0^{a,b}=0$ whenever $a$ or $b$ are odd.
The values of $g_0^{n,0}$ for $n$ even remain free, but must satisfy a
condition following from the next adiabatic order $g_1^{a,b}$.

For the first adiabatic order, we now consider $g_0^{a,b}$ weakly
time-dependent via $\langle\hat{q}\rangle$, but assume
$g^{a,b}-g_0^{a,b}$ time independent. Solutions to the resulting
equations $\{g_1^{a,b}-g_0^{a,b},H_Q\}=0$ provide the first adiabatic
order\footnote{The $n$-th adiabatic order $g_n^{a,b}$ is assumed to
  satisfy $\{g_n^{a,b}-g_{n-1}^{a,b},H_Q\}=0$ and thus solves
  $\{g_n^{a,b},H_Q\}= \dot{g}_{n-1}^{a,b}$. Iterating over $n$, this
  provides algebraic equations for all orders.} $g_1^{a,b}$, whose time
dependence is given by derivatives of the zeroth (adiabatic) order
moments:
\[
 \{g_1^{a,b},H_Q\}= \omega\left(ag_1^{a-1,b+1}-
b\left(1+\frac{U''(\langle\hat{q}\rangle)}{m\omega^2}\right)
 g_1^{a+1,b-1}\right)=\dot{g}_0^{a,b}\,.
\]
This equation implies
\[
 \sum_{b\:{\rm even}} {(a+b)/2 \choose b/2}
\left(1+\frac{U''(\langle\hat{q}\rangle)}{m\omega^2}\right)^{b/2}\dot{g}_0^{a,b}=0
\]
and requires
$g^{n,0}_0=C_n(1+U''(\langle\hat{q}\rangle)/m\omega^2)^{-n/4}$.  The
remaining constant $C_n$ can be fixed by requiring the moments to be
of harmonic-oscillator form for $U=0$: $C_n=2^{-n}n!/(n/2)!$.  In
particular, $g^{qq}_0= \frac{1}{2}
(1+U''(\langle\hat{q}\rangle)/m\omega^2)^{-1/2}$, and the quantum
correction to the effective force is $-\frac{1}{4}(\hbar/m\omega)
U'''(\langle\hat{q}\rangle)(1+U''(\langle\hat{q}\rangle)/m\omega^2)^{-1/2}$
as it arises from an effective potential
\begin{equation} \label{Veff}
V_{\rm eff}(q)= \frac{1}{2}m\omega^2q^2+ U(q)+
\frac{1}{2}\hbar\omega\sqrt{1+\frac{U''(q)}{m\omega^2}}
\end{equation}
from $\frac{1}{2}\hbar\omega (g_0^{qq}+g_0^{pp})+
\frac{1}{2}(\hbar/m\omega) U''(\langle\hat{q}\rangle) g_0^{qq}$ in
(\ref{HQexpand}).  This function agrees with path-integral
calculations of the low-energy effective action
as derived by \cite{EffAcQM}. Here, only the
zeroth adiabatic order combined with first order in $\hbar$ has been
used. To second order computed by \cite{EffAc}, there is also a
correction for the mass term, still in agreement with \cite{EffAcQM}.

The adiabatic approximation is an example for situations in which the
coupled quantum evolution of moments can be further reduced to result in
explicit effective forces. While the forces come from coupling terms
between expectation values and quantum variables such as fluctuations,
the latter do not appear explicitly. Quantum effects then manifest
themselves indirectly in the form of effective terms depending on the
expectation values, their quantum origin being indicated only by the
presence of $\hbar$ but not by explicit quantum degrees of
freedom. Such effective terms cannot always be derived since some
regimes, where effective descriptions may well apply, do require the
larger freedom of higher-dimensional effective systems coupling
quantum degrees of freedom explicitly. For low-energy effective
potentials, the adiabatic approximation is responsible for the
reduction to a system of classical form as far as degrees of freedom
are concerned.

From the derivation we can also see how the state dependence is a
crucial part of effective equations, even though the final expression
(\ref{Veff}) for the potential seems state independent. To fix all
free constants in effective equations for anharmonic systems, in
particular $C_2$, we had to refer to the harmonic-oscillator ground
state.  We are thus expanding around the vacuum of the free, solvable
theory whose moments are known. Quantum corrections from the
interacting vacuum have been derived in passing, by obtaining the
leading adiabatic orders of $g^{qq}$. For the low-energy effective
potential, this result was re-inserted in the equations of motion,
somewhat hiding the state dependence.

With the higher-dimensional viewpoint of effective systems, keeping
some of the moments as independent parameters, we obtain extra
information about the interacting theory not seen in a simple
effective potential. For instance, from our intermediate calculations
we directly have $g_0^{qq}=\frac{1}{2}
(1+U''(\langle\hat{q}\rangle)/m\omega^2)^{-1/2}$, while $g_0^{pp}=
(1+U''(\langle\hat{q}\rangle)/m\omega^2) g_0^{qq}=\frac{1}{2}
(1+U''(\langle\hat{q}\rangle)/m\omega^2)^{1/2}$ and $g_0^{qp}=0$. To
zeroth adiabatic order, the interacting ground state keeps saturating
the uncertainty relation, but fluctuations are no longer exactly
constant.

\subsection{Quantum cosmology}

In relativistic systems, there is no absolute time with evolution
generated by a Hamiltonian. Rather, relativistic systems are subject
to a Hamiltonian constraint $C$. It generates arbitrary changes of the
time coordinate as gauge transformations $\delta_{\epsilon}f= \epsilon
\{f,C\}$ for phase-space functions $f$. From observable quantities $O$
left unchanged by gauge transformations, that is
$\{O,C\}=0$, dynamical properties follow. Since the invariance
condition $\{O,C\}=0$ removes one dimension from the initial phase
space, for consistency we must require $C=0$ as a
constraint.\footnote{Factoring out the Hamiltonian flow generated by
  the constraint $C$ via its Hamiltonian vector field
  $X_C=\{\cdot,C\}$, we obtain a projection $\pi\colon M \to M/X_C$
  from the original phase space $M$ by identifying all points along
  the orbits of $X_C$. All observables $O$ naturally descend to the
  factor space since they are constant along the orbits, and so does
  $C$. In this way, we obtain a complete set of functions on $M/X_C$.
  On the factor space, we have a natural Poisson structure
  $\{f,g\}_{M/X_C}= \{\pi^*f,\pi^*g\}_M$, pushing forward the Poisson
  bivector via $\pi$. This Poisson structure is degenerate:
  $\{O,C\}_{M/X_C}=0$ for all functions $O$ on $M/X_C$. The constraint
  $C$ becomes a Casimir function on the factor space, and symplectic
  leaves of the Poisson structure are given by $C={\rm const}$. Any
  leaf carries a non-degenerate symplectic structure and can be taken
  as a reduced phase space, but $C=0$ is distinguished: In this case,
  we can write gauge transformations as $\delta_{\epsilon}
  f=\{f,\epsilon C\}$ even for phase-space functions $\epsilon$. More
  details of Poisson geometry in the context of constrained systems are
  described by \cite{brackets}.}

Constrained formulations can be introduced also for non-relativistic
systems by parameterization, adding a time degree of freedom $t$ with
momentum $p_t$ and replacing the Hamiltonian $H$ by the Hamiltonian
constraint $C=p_t-H$. For time-independent Hamiltonians, $p_t$ is
gauge invariant while the gauge transformation of $t$ is
$\delta_{\epsilon}t=\epsilon \{t,C\}= \epsilon$; $t$ can be changed at
will. For the remaining observables,
\[
 0=\{O,C\}= \frac{{\rm d}O}{{\rm d}t}-\{O,H\}
\]
imposes Hamilton's equations of motion.

For relativistic quantum systems, the effective techniques described
so far cannot directly be applied. There are extensions to effective
constraints briefly described later, as developed by
\cite{EffCons,EffConsRel}. But more simply, effective equation
techniques can be used if systems are first deparameterized, reverting
the above procedure. If there is a
variable $\phi$, then called internal time, such that the Hamiltonian
constraint can be written as $C=p_{\phi}-H(q,p)$ with the momentum
$p_{\phi}$ of $\phi$ and a function $H$ independent of $\phi$ and
$p_{\phi}$, gauge transformations generated by the constraint take the
form
\[
 \delta_{\epsilon}f=\epsilon \{f(q,p,\phi,p_{\phi}),C\}= 
\epsilon \left(\frac{\partial f}{\partial \phi}- \{f,H\}\right)\,.
\]
Gauge invariant quantities of the theory are thus those evolving in
the usual Hamiltonian way as generated by the $\phi$-Hamiltonian $H$.

Once deparameterized, the observables of a constrained system can be
derived by analyzing an ordinary Hamiltonian flow.  At this stage,
effective techniques as described before can be applied to
quantizations of deparameterized models. Effective equations of motion
derived from
\begin{equation}\label{dOdphi}
 \frac{{\rm d}\langle\hat{O}\rangle}{{\rm d}\phi} =
 \frac{\langle[\hat{O},\hat{H}]\rangle}{i\hbar}
\end{equation}
then provide means to solve for quantum observables
$\langle\hat{O}\rangle(\phi)$ and their physical evolution. For the
initial constrained system, solutions $\langle\hat{O}\rangle(\phi)$
are observables as functions on the full phase space including
$(\phi,p_{\phi})$. (Remaining phase space variables enter the
expression $\langle\hat{O}\rangle(\phi)$ via initial values taken for
them when solving the differential equations (\ref{dOdphi}).) The
deperameterization endows $\langle\hat{O}\rangle(\phi)$ with the
interpretation as an observable $\langle\hat{O}\rangle$ evolving with
respect to the internal time $\phi$. This is the relational picture
for interpreting constrained dynamics, developed classically by
\cite{BergmannTime}, \cite{GeomObs1,GeomObs2} and
\cite{PartialCompleteObs,PartialCompleteObsII}.

Examples for such systems in cosmology are homogeneous models sourced
by a free, massless scalar. Its energy density is purely kinetic,
$\rho=\frac{1}{2}a^{-6}p_{\phi}^2$, such that the Friedmann equation,
solved for $p_{\phi}$, provides a Hamiltonian for $\phi$-evolution.
For a spatially flat Friedmann--Robertson--Walker model, the
$\phi$-Hamiltonian then turns out to be quadratic in suitable canonical
variables: for instance, the Hubble parameter ${\cal H}=\dot{a}/a$ is
canonically conjugate to the volume $V=a^3$, and the Friedmann
equation
\[
 \left(\frac{\dot{a}}{a}\right)^2= \frac{8\pi G}{3}\rho= \frac{4\pi
   G}{3}\frac{p_{\phi}^2}{a^6}
\]
tells us that $p_{\phi}^2\propto V^2{\cal H}^2$.  Upon taking a square
root, we have a quadratic $\phi$-Hamiltonian.  Based on this
observation, state properties have been determined by
\cite{BouncePert,BounceCohStates}.  (Strictly speaking, the
$\phi$-Hamiltonian is of the form $|qp|$ which is not quadratic.
However, for effective equations one can show that the absolute value
can be dropped, providing a linear quantum system. We only have to
require an initial state to be supported on a definite part of the
spectrum of $\widehat{qp}$, either the positive or the negative one,
which is then preserved in time since $\hat{H}$ is preserved. The
absolute value just amounts to multiplication with $\pm1$. For
initially semiclassical states, this requirement does not lead to
restrictions for low-order moments.)

For other systems, perturbation theory can be used as described above
for anharmonic oscillators. A crucial difference, according to the
analysis by \cite{BouncePot}, is that no adiabatic regime has so far
been found for quantum cosmology, blocking the complete expression of
quantum variables in terms of an effective potential. On the other
hand, higher-dimensional effective systems, where quantum variables are
taken as independent variables subject to their own evolution, can be
analyzed and show how states back-react on the expectation-value
trajectories.

If we include a mass term or a potential for the scalar, the system
becomes ``time-dependent'' in $\phi$. Extra care is required, but
perturbation theory still applies for small and flat potentials. The
justification for this procedure in the time-dependent case comes from
an extension of the effective-equation procedure to constrained
systems introduced by
\cite{EffCons,EffConsRel} without requiring deparameterization.
Quantum constraint operators then imply the existence of infinitely
many constraints $\langle \widehat{\rm pol}\hat{C}\rangle$ on the
quantum phase space, in general all independent for different
polynomials $\widehat{\rm pol}$ in the basic operators. This large
number of constraints restricts not only expectation values (as the
classical analog) but also the corresponding quantum variables. A
complete reduction to the physical state space results, without
requiring a deparameterization to exist.  With these general
techniques, effective equations for relativistic systems are thus
fully justified. As discussed in Sec.~\ref{s:Cons}, inhomogeneous
models bring in a new level due to the anomaly problem, which does not
arise for a single classical constraint.

\subsection{Symplectic structure}

Quantum corrections in canonical effective equations come from
corrected Hamiltonians or corrected constraints as expectation values
of Hamiltonian (constraint) operators.  The Hamiltonian $H$ or
constraints $C_I$, together with Lagrange multipliers $N^I$ (for
gravity lapse and shift multiplying the Hamiltonian and diffeomorphism
constraints), form an important contribution to the action. But the
action
\[
 S[q(t),p(t); N^I]= \int{\rm d}t\left(\dot{q}p-H(q,p)-N^IC_I(q,p)\right)
\]
of a Hamiltonian system in canonical form has an extra contribution,
the one that determines the symplectic structure between configuration
and momentum variables.  One might wonder whether it is enough to look
for quantum corrections in the Hamiltonian or the constraints without
correcting the symplectic structure.  With symplectic-structure
corrections, an effective action might take a different form than
suggested by an analysis only of Hamiltonians and constraints.

There might indeed be corrections to the symplectic structure, but
they would follow from the same algebraic notions used for canonical
effective equations. Poisson brackets of quantum variables are given
by expectation values of commutators as used before; see (\ref{AB}).
Any potential corrections to the symplectic structure have thus been
taken care of by applying that formula consistently.  If there are no
changes in Poisson-bracket relations for effective equations, it is
only because they do not change for expectation values of basic
operators: they have commutator relations mimicking the classical
Poisson algebra, which is linear for the basic objects.  Taking
expectation values of these linear structures does not lead to
corrections. For canonical basic operators $\hat{q}$ and $\hat{p}$,
for instance, we have
$\{\langle\hat{q}\rangle,\langle\hat{p}\rangle\}=
\langle[\hat{q},\hat{p}]\rangle/i\hbar=1$.

The Poisson structure changes only because the dimension of the phase
space increases by the quantum variables $G^{a,b}$. These variables
satisfy Poisson relationships following from the quantum theory, but
this does not affect the Poisson brackets of expectation values of
basic operators. In an action, the symplectic term for classical
variables remains unchanged. But if a higher-dimensional effective
system with independent quantum variables is used, their symplectic
terms add to the action. The symplectic structure is only extended to
include new degrees of freedom; it is not quantum corrected.

Sometimes, one can make assumptions\footnote{A quantum cosmological
  model with assumptions for semiclassical states has been analyzed by
  \cite{Taveras}.} about the dependence of quantum variables on
expectation values, or even derive those by an adiabatic
approximation. If this is done, one can insert such expressions,
schematically $G^I(q,p)$, into the symplectic form for quantum
variables, of the form $\Omega_{IJ}(G) {\rm d}G^I\wedge {\rm d}G^J$
where $\Omega_{IJ}(G)$ follows from (\ref{AB}) purely kinematically.
Then, a term $2\Omega_{IJ}(\partial G^I/\partial q) (\partial
G^J/\partial p) {\rm d}q\wedge {\rm d}p$ results which would add to
the symplectic term of expectation values.  Now, the correction has
dynamical information via the partial solutions $G^I(q,p)$. (For the
specific example of anharmonic oscillators, no such corrections arise
since the effective equations for $g^{ab}$, and thus their solutions,
depend only on $\langle\hat{q}\rangle$, not $\langle\hat{p}\rangle$.)
With independent quantum variables, as they are most often required,
no symplectic-structure corrections result. Expectation values of
Hamiltonians or constraints are then the key source for quantum
corrections.

For quantum gravity, we expect higher-curvature terms in an effective
action, and thus higher-derivative terms.  By the preceding
discussion, this can only come from independent moments $G^{a,b}$.
But quantum gravity has additional implications, such as the emergence
of discrete spatial structures expected in particular from loop
quantum gravity. They, too, must affect the effective dynamics.

\section{Discrete dynamics}

Our main interest from now on will be to apply effective equation
techniques to canonical quantum gravity. A strict derivation requires
detailed knowledge of the mathematical properties of operators
involved, as well as information about the representation of states.
Effective equations and effective constraints are, after all, obtained
via expectation values of Hamilonians or Hamiltonian constraint
operators. These operators are first constructed within the full
quantum theory of gravity or one of its models, and the resulting
constructs will be sensitive to properties of the basic operators,
analogous to $\hat{q}$ and $\hat{p}$ in quantum mechanics. Some of
these properties will descend to the effective level once expectation
values of Hamiltonians are computed, and then affect also the
effective dynamics. At this stage, general effective techniques
naturally tie in with specific constructions of a concrete quantum
theory at hand.

\subsection{Loop quantum gravity}

Classically, the canonical structure of general relativity is given by
the spatial part $q_{ab}$ of the space-time metric $g_{ab}$, as well
as momenta related to its change in time, or the extrinsic curvature
$K_{ab}$ of the spatial slices $\Sigma$. These specific quantities are
meaningful only when a choice for time, a time function $t$ such that
$\Sigma\colon t={\rm const}$, has been made, and so this formalism is
often seen as breaking covariance. But space-time covariance is broken
only superficially and is restored when all the dynamical constraints
have been solved --- in addition to a local Hamiltonian constraint,
three components of the diffeomorphism constraint to generate all four
independent space-time coordinate changes. The theory is, after all,
equivalent to general relativity in its Lagrangian form; just
formulating it in different variables cannot destroy underlying
symmetries. In fact, the Hamiltonian formulation still shows the full
generators of all gauge transformations in explicit form, by the
constraints it implies for the fields. By the general theory of
constraint analysis, a discussion of gauge at the Hamiltonian
level\footnote{For an introduction to these techniques see
\cite{CUP}.} is then even more powerful than at the Lagrangian one.

\subsubsection{Smearing}

Before addressing quantum constraints, the quantum theory must be set
up. For a well-defined quantization one turns the basic fields into
operators such that the classical Poisson bracket is reflected in
commutator relationships. In constructions of quantum field theories
one has to face the problem that Poisson brackets of the fields
involve delta functions since they are non-vanishing only when the
values of two conjugate fields are taken at the same point, as in
$\{\phi(x),p_{\phi}(y)\}=\delta(x,y)$ for a scalar field. A simple but
powerful remedy is to ``smear'' the fields by integrating them against
test functions over space. Again for a scalar field, we could use
$\phi[\mu]:=\int{\rm d}^3x \sqrt{\det q} \mu(x)\phi(x)$ for which
$\{\phi[\mu],p_{\phi}(y)\}= \int{\rm d}^3x \mu(x)\delta(x,y)=\mu(y)$.
The Poisson algebra for the smeared fields is free of delta functions
thanks to the integration. Moreover, for sufficiently general classes
of test functions $\mu(x)$, smeared fields capture the full
information contained in the local fields and can be used to set up a
general theory.

After smearing, a well-defined Poisson algebra of basic objects
results, ready to be turned into an operator algebra by investigating
its representations. For gravity, we would use smeared versions of the
spatial metric and its change in time. But here, a second problem
arises. One of the dynamical fields to be quantized is the spatial
metric, but to smear fields we need a metric for the integration
measure. This is no problem when fields to be quantized are
non-gravitational. As with the scalar in the example above, we would
use the background space-time on which the scalar moves, obtaining
quantum field theory on a given, possibly curved background. But what
do we do if the metric itself is one of the fields to be quantized? If
we use it to smear itself, the resulting object becomes ugly,
non-linear and too complicated as one of the basic quantities of a
quantum field theory. If we introduce a separate metric just for the
purpose of smearing the dynamical fields of gravity, this auxiliary
input is likely to remain in the results derived from the theory. We
would be quantizing gravitational excitations on a given space-time
background, not the full gravitational field or space-time itself.
We would be violating the great insight of general relativity by
formulating physics on an auxiliary space-time, rather than realizing
gravity as the manifestation of space-time geometry.

Fortunately, it is possible to smear fields and yet avoid the
introduction of auxiliary metrics. Back to the scalar, we can choose
to smear $p_{\phi}$ instead of $\phi$. The momentum of a scalar field
is a scalar density; it transforms under coordinate changes with an
extra factor of the Jacobian for the coordinate transformation. This
is a direct consequence of the definition $p_{\phi}=\partial{\cal
L}/\partial\dot{\phi}$ as a derivative of the Lagrangian density.
Explicitly, the canonical variable $p_{\phi}=\sqrt{\det q}\dot{\phi}$
already carries the correct measure factor which need not be
introduced by an auxiliary metric.\footnote{Although $\sqrt{\det q}$
appears in the relationship between $p_{\phi}$ and $\dot{\phi}$, from
the viewpoint of Poisson geometry $p_{\phi}$ (but not $\dot{\phi}$) is
independent of the metric: $\{p_{\phi},p^{ab}\}=0$ for the momenta
$p^{ab}$ of $q_{ab}$.} The smeared version
$p_{\phi}[\lambda]:=\int{\rm d}^3y\lambda(y)p_{\phi}(y)$ is
well-defined for any function $\lambda$, and it suffices to remove
delta functions from the Poisson algebra:
$\{\phi(x),p_{\phi}[\lambda]\}= \lambda(x)$.

\subsubsection{Holonomies and fluxes}

For tensorial fields as we have them in gravity,
background-independent smearings are often more difficult to
find. Loop quantum gravity has provided suitable procedures for
general relativity, but for this it must first transform from metric
variables to connections with their conjugates, densitized
triads. Connections and densitized vector fields turn out to have just
the right transformation properties under coordinate changes that they
can, with one loopy trick, be smeared background independently. A
well-defined quantization results with several immediate implications
for the basic operators encoding spatial geometry, as well as
far-reaching and sometimes surprising consequences in the resulting
dynamics.

Instead of using the spatial metric $q_{ab}$, spatial geometry is
expressed by a densitized triad $E^a_i=\sqrt{\det q} e^a_i$ such that
$E^a_iE^b_i= \det q \, q^{ab}$. The densitized triad is canonically
conjugate to $K_a^i:=K_{ab}e^{bi}$ in terms of extrinsic curvature
$K_{ab}$. To obtain a connection with its useful transformation
properties, we finally follow \cite{AshVar} and \cite{AshVarReell} to
introduce the Ashtekar--Barbero connection $A_a^i=\Gamma_a^i+\gamma
K_a^i$ with the spin connection $\Gamma_a^i$ compatible with the
densitized triad and a positive real number $\gamma$, the
Barbero--Immirzi parameter (whose role for quantum geometry was
realized by \cite{Immirzi}).

The elementary objects loop quantum gravity takes for its
representation are holonomies and fluxes,
\[
 h_e(A)={\cal P}\exp\left(\int_e{\rm d}s
    \dot{e}^aA_a^i\tau_i\right)\quad\mbox{ and }\quad F_S(E)=
\int_S{\rm d}^2y n_aE^a_i\tau_i\,,
\]
first used in this context by \cite{LoopRep}.  A key advantage is that
their algebra under Poisson brackets is well-defined, free of
delta-functions (unlike the algebra of fields), and yet independent of
any background metric. Only the dynamical fiels $A_a^i$ and $E^a_i$
are used, together with kinematical objects such as curves
$e\subset\Sigma$ and surfaces $S\subset\Sigma$ as well as the tangent
vectors $\dot{e}^a$ and co-normals $n_a$ to them, but no independent
metric structure.\footnote{The co-normal, unlike the normal $n^a$, is
  metric independent: for a surface $S\colon f={\rm const}$,
  $n_a=({\rm d}f)_a$.} All spatial geometrical properties are
reconstructed from $E^a_i$ via fluxes, and space-time geometry follows
with $A_a^i$ via holonomies once equations of motion (or rather the
constraints of relativity) are imposed. In this way, loop quantum
gravity provides a framework for background-independent quantum
theories of gravity.

Once a well-defined algebra of basic objects has been chosen, one can
determine its representations to arrive at possible quantum theories.
In the connection representation, a complete set of states
$\psi(A_a^i)$ is generated by holonomies as multiplication operators
acting on $\psi(A_a^i)=1$.  In the case of loop quantum gravity, this
has an immediate and general consequence. Holonomies take values in
SU(2), and fluxes, depending on the momenta $E^a_i$, become derivative
operators on SU(2) --- just like angular momentum in quantum
mechanics. For a dense set of states, only a finite number of
holonomies (along curves intersecting the flux surface $S$) contribute
to a given flux. Finite sums of angular momentum operators with
discrete spectra have a discrete spectrum, too: spatial geometry is
discrete; flux operators quantizing the densitized triad and thus
encoding spatial geometry acquire discrete spectra. So do spatial
geometrical quantities such as areas and volumes as constructed by
\cite{AreaVol,Area,Vol2}. No extra assumptions are required; one
merely has to fix the basic algebra and follow mathematical procedures
to analyze its representations. A different algebra might lead to
other properties, possibly not with discrete spatial geometry. But no
alternative procedure providing a well-defined and smeared, yet
background independent quantization has been found. The holonomy-flux
algebra suggested by its natural smearing, on the other hand, has a
unique irreducible, cyclic, diffeomorphism covariant representation as
proven by \cite{LOST} and\cite{WeylRep}. Most of these properties are
described by \cite{SahlmannCT}.

\subsubsection{Dynamics}

Kinematical properties are elegant, simple, and largely unique. The
theory starts to get considerably more messy when its dynamics is
considered. Here, two major tasks must be performed: Dynamical
operators, mainly the Hamiltonian constraint, must be defined from the
basic ones, holonomies and fluxes. This is the constructive part of
the task. For suitable constructions of the Hamiltonian constraint,
the dynamics of the theory must then be evaluated, a process still
full of several open issues.

At the constructive stage, many choices can be made, and strong
consistency conditions must be respected. We are witnessing an epic
battle between the liberating anarchy of choice and the uniformizing
tyranny of constraints. Just writing a Hamiltonian constraint operator
is tedious but possible in many ways. There are ubiquitous factor
ordering ambiguities, as well as other choices specific to loop
quantum gravity. On the other hand, the Hamiltonian constraint
provides not only an equation of motion, it also generates a crucial
part of the gauge transformations responsible for general covariance.
A consistent quantization must keep gauge degrees of freedom as gauge,
and not overly restrict the number of physical, non-gauge degrees of
freedom. All this can usefully be formalized, as we will see later,
providing strong algebraic conditions. They are so strong that to
date, despite the abundant freedom of choices in constructing
Hamiltonian constraint operators, no full consistent version has been
found.

Once a consistent version of the dynamics exists, it must be
evaluated. We must find solutions, and determine the physical
observables they provide. Their values, finally, can be used for
predictions such as small deviations from the expected classical
behavior. Since no full consistent version has been found yet, and
since all potential condidates are highly complicated, construction
issues of the dynamics have so far dominated strongly over evaluation
issues. In several model systems, on the other hand, the dynamics can
often be simplified so much that it can be analyzed rather
explicitly. Many useful techniques are now available, mainly in the
context of effective equations. We will come back to these
applications after discussing more details of the construction side of
the problem.

\subsubsection{Hamiltonian constraint}

Specifically, the Hamiltonian constraint of general relativity in
Ashtekar variables is
\begin{eqnarray*}
 C[N] &=& \int_{\Sigma} \mathrm{d}^3x N(x)
 \left(\epsilon_{ijk}F_{ab}^i \frac{E^a_jE^b_k}{\sqrt{|\det
E|}} \right.
  -\left.
2(1+\gamma^{-2})
 K_a^iK_b^j\frac{E^{[a}_iE^{b]}_j}{\sqrt{|\det E|}}
 \right)
\end{eqnarray*}
with the curvature $F_{ab}^i$ of the Ashtekar connection and extrinsic
curvature $K_a^i$. It has to vanish for all lapse functions $N(x)$,
thus providing infinitely many constraints. If $C[N]$ is to be turned
into an operator, using the basic expressions for holonomies and
fluxes, several obstacles must be overcome.

First, there is the potentially singular inverse determinant of the
densitized triad. No direct quantization exists since the densitized
triad has been quantized to flux operators with discrete spectra,
containing zero. Such operators lack densely defined inverses.
Nevertheless, quantizations with the appropriate inverse as the
semiclassical limit can be obtained making use of the classical
identity
\begin{equation}\label{InvTriad}
 \left\{A_a^i,\int{\sqrt{|\det E|}}\mathrm{d}^3x\right\}\propto
 \epsilon^{ijk}\epsilon_{abc} \frac{E^b_jE^c_k}{{\sqrt{|\det E|}}}
\end{equation}
(or variants) as introduced by \cite{QSDI}.  There is no inverse on
the left-hand side. Instead, the expression involving the densitized
triad is the spatial volume which (with some regularization) can
directly be quantized. The connection components can be expressed via
holonomies, and the Poisson bracket will, at the quantum level, be
quantized to a commutator divided by $i\hbar$. A well-defined operator
results with the right-hand side of (\ref{InvTriad}) as the desired
semiclassical limit by construction.

The remaining factors in the Hamiltonian constraint involve the
Ashtekar curvature as well as extrinsic curvature. For the curvature
components $F_{ab}^i$, we can use
\begin{equation}\label{F}
 s_1^as_2^bF_{ab}^i(x)\tau_i=
\Delta^{-1}(h_{\lambda}-1) +O(\Delta)
\end{equation}
where $\lambda$ is a small loop starting at a point $x$, spanning a
coordinate area $\Delta$, and with tangent vectors $s_1^a$ and $s_2^a$
at $x$. On the right-hand side, the holonomy $h_{\lambda}$ can readily
be quantized, and to leading order provides curvature components as
required for the constraint.

Extrinsic curvature, finally, is a more complicated object in terms of
the basic ones but can be obtained from what has been provided so far:
\[
K_a^i\propto \left\{A_a^i,\left\{\int{\rm d}^3x F_{ab}^i 
\frac{\epsilon^{ijk}E^a_jE^b_k}{\sqrt{|\det
E|}},\int{\sqrt{|\det E|}}\mathrm{d}^3x\right\}\right\}
\]
expresses extrinsic curvature in terms of a nested Poisson bracket
involving the spatial volume and the first term of the constraint,
already provided by the preceding steps.

\begin{figure}
\begin{center}
\includegraphics[width=14cm]{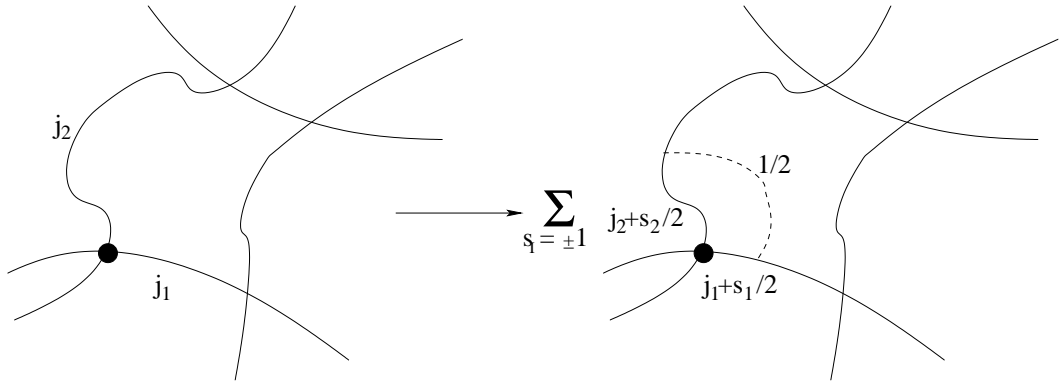}
\end{center}
\caption{Schematically, the local action of the Hamiltonian constraint
  operator on a state. States are generated by holonomies as
  multiplication operators, visualized by the graph formed by all
  curves $e$ used. Moreover, labels $j_e$ determine matrix representations of
  SU(2)-valued $h_e(A)$. Due to (\ref{F}), new curves and vertices are
  typically created when the Hamiltonian constraint acts. \label{f:Local}}
\end{figure}

In this way, holonomy and flux operators make up the Hamiltonian
constraint operator $\hat{C}$ as a densely defined operator
(including, in the non-vacuum case, regular matter Hamiltonians again
exploiting (\ref{InvTriad}) following \cite{QSDV}). It determines the
physical solution space by its kernel: physical states $\psi(A)$,
assumed again in the connection representation, must satisfy
$\hat{C}\psi(A)=0$, or $(\hat{C}+8\pi G\hat{H}_{\rm
  matter})\psi(A,\phi,\ldots)=0$ if matter is present. The action of
the constraint is rather complicated, as visualized schematically in
Fig.~\ref{f:Local}

Physical states of interest normally belong to zero in the continuous
part of the spectrum of $\hat{C}$, which requires the introduction of
a new physical Hilbert space spanned by the solutions to the quantum
constraint equation and equipped with a suitable physical inner
product. Constructing the physical Hilbert space leads over to
evaluating the dynamics, for physical states would provide predictions
from expectation values of observables. But this stage has been
brought to completion only in a few simple (and very special) models;
in general the outlook toward a full implementation is rather
pessimistic.

Here, effective constraint techniques
become useful because they allow one to address physical properties,
corresponding to observables in the physical Hilbert space without
having to deal explicitly with states. Conditions from the physical
inner product are rather implemented by reality conditions for
expectation values and quantum variables such as fluctuations, which
can be done more simply and more generally than for entire states.
Effective constraints thus provide a good handle on generic properties
of physical observables, at least in semiclassical regimes where not
all the moments need be considered. Especially in cosmological
situations they provide an ideal framework. In inhomogeneous contexts,
they allow a detailed discussion of the anomaly
issue, and show whether the different effects expected from the basic
operators of quantum gravity can lead to a consistent form of the
dynamics.

Consistency conditions in the presence of ambiguities are useful, for
they constrain the choices. But the question remains whether loop
quantum gravity can be fully consistent at all --- implementing
covariance at the quantum level might, after all, not leave any
consistent physical states. If the consistency conditions are weak, on
the other hand, ambiguities would remain even at the physical level.
In between, at a razor-thin balance between ambiguity and constraints,
lies the case of a unique consistent theory, a possibility which
exists in loop quantum gravity but for which at present no evidence
has been found.

In model systems it has at least been shown that consistency in the
form of covariance can be achieved even in the presence of quantum
corrections resulting from the discreteness. This statement may come
as a surprise, for discrete spatial or space-time structures are
naively expected to break local Lorentz or even rotational symmetries.
Nevertheless, at the level of effective equations one can see that
covariance can be respected --- but it cannot leave the classical
algebra of local symmetries invariant. While quantum-gravity
corrections provide a consistent deformation of general relativity ---
preserving the number of independent gauge generators --- the algebra
is truly deformed. Gauge transformations are no longer local Lorentz
or coordinate transformations, but of a different type relevant for
the quantum space-time structures derived from loop quantum gravity.

Much of these conclusions makes use of results obtained through
several years for models of loop quantum cosmology. We will now start
a general exposition beginning with the simplest isotropic models,
introducing the key features (and sources of ambiguities), and then
leading over to the discussion of covariance in the next section.

\subsection{Isotropic loop quantum cosmology}

Loop quantum cosmology provides quantizations of symmetry-reduced
models of general relativity, starting with isotropic ones in the
simplest cases. It is thus a minisuperspace quantization, but not only
that. There are explicit relationships between the states and the
algebra of basic operators in a model of loop quantum cosmology, which
shows how they descend from the analogous expressions in the full
theory. For different kinds of reductions 
such as implementing isotropy, homogeneity or spherical symmetry, this
has been constructed by \cite{AnisoPert,SphSymm,InhomLattice}. In
particular, properties of the basic operators, for instance the
discreteness of the flux spectra, are preserved and realized also in
cosmological models.  Crucial implications for the discreteness of
spatial geometry can then be analyzed in their dynamical context.
Qualitatively, all applications of loop quantum cosmology rely on this
preservation of discreteness properties by the symmetry reduction
introduced by \cite{SymmRed}.

While existing methods do not allow one to derive the dynamics of
models directly from full constraint operators, it can be constructed
in an analogous way using all the construction steps sketched for the
full constraint. Also here, crucial properties are preserved. And even
though the incompletely known relationship to the full theory
introduces additional ambiguities in the dynamics of models, reliable
conclusions can still be drawn provided the considerations are
generic enough. Here, the need for sufficiently general
parameterizations of ambiguities arises.

\subsubsection{Basic variables}

In spatially flat isotropic models, the basic canonical fields reduce
to $A_a^i=\tilde{c} \delta_a^i$ and $E^a_i=\tilde{p}\delta^a_i$ with
two dynamical variables $\tilde{c}$ and $\tilde{p}$. Relating the
densitized triad to the spatial metric shows that $|\tilde{p}|=a^2$ is
given by the scale factor. (Due to the freedom of choosing an
orientation of the triad, $\tilde{p}$ can take both signs.)
Classically, $\tilde{c}=\gamma \dot{a}$ for spatially flat models. 

To define the Poisson structure, we pull back the full symplectic form
$(8\pi\gamma G)^{-1}\int{\rm d}^3x \delta A_a^i\wedge\delta E^a_i$ by
the embedding $(\tilde{c},\tilde{p})\mapsto (A_a^i,E^a_i)=(\tilde{c}
\delta_a^i,\tilde{p}\delta^a_i)$. Due to homogeneity, the resulting
integral diverges if we integrate over all of space and space is
infinite, but homogeneity also implies that we may just integrate over
any finite chunk of coordinate volume $V_0$, say, and still get the
complete reduced symplectic form, $3V_0(8\pi\gamma G)^{-1} {\rm
  d}\tilde{c}\wedge {\rm d}\tilde{p}$. It depends on the arbitrary
$V_0$, which can be hidden in the canonical variables by redefining
them as
\[
 c=V_0^{1/3}\tilde{c} \quad\mbox{and}\quad p=V_0^{2/3}\tilde{p}\,.
\]
In the end, physical quantities must be ensured to depend only on
combinations of $c$, $p$ and possibly other ingredients such that they
are insensitive to changes of $V_0$. (Changing $V_0$ is independent of
changing coordinates, thereby rescaling the scale factor. While
$\tilde{c}$ and $\tilde{p}$ depend on the scaling but not on $V_0$,
$c$ and $p$ depend on $V_0$ but not on the scaling.)

Isotropic  connections and densitized triads result in
specific versions of holonomies and fluxes. They take especially simple
forms when evaluated along curves or surfaces making use of structures
provided by the homogeneity group, i.e.\ curves along generators of
translations and surfaces transversal to them. In a homogeneous
context, fluxes are simply the triad components multiplied with the
coordinate area of the surface, and holonomies can be computed
explicitly: $h_{e_j}(A_a^i)= \exp(\mu c \tau_j)$ with $\mu=
\ell_0/V_0^{1/3}$ for a curve $e_j$ of coordinate length $\ell_0$
along the direction $\dot{e}_j^a=(\partial/\partial x^j)^a$ in
Cartesian coordinates.  If we take surfaces for fluxes of edge length
the same size as $e_j$, fluxes are $F_{S_j}(E^a_i)= \mu^2 p \tau_j$
for a surface $S_j$ transversal to $(\partial/\partial x^j)^a$.

These curves and surfaces are particularly useful because they can be
arranged in a regular lattice of spacing $\ell_0$. While such a
setting is not the most general one (and other curves would lead to
different expressions for holonomies as computed for instance by
\cite{AinvinA}), it allows one to capture all the degrees of freedom
in a homogeneous model. More importantly, it allows for rough contact
with the full theory, for the regular lattices envisioned here can be
thought of as lattices for states in the full theory. In particular
for comparisons of actions of the dynamical operators, such a
relationship is convenient. It also allows us to find a useful
interpretation of the parameter $\mu$ characterizing isotropic
holonomies and fluxes: For a regular lattice of spacing $\ell_0$,
contained in a region of size $V_0$, $\mu^3=\ell_0^3/V_0=: {\cal
  N}^{-1}$ is the inverse number of lattice sites. This parameter can
be seen as providing information about an underlying discrete
inhomogeneous state which appears even in a
homogeneous reduction. In fact, properties of ${\cal N}$ such as its
size or possible dynamical features play a crucial role in Hamiltonian
quantum evolution.

\subsubsection{Representation}

From isotropic holonomies and fluxes, we can construct their loop
representation by analogy with the full theory. States in the
connection representation are functionals of holonomies and are thus
superpositions of the basic states $|\mu\rangle$, or $\exp(i\mu c)$ as
functions of the isotropic connection component. Here, $\mu$ is an
arbitary real parameter. An inner product of states can be derived
from integration theory on spaces of connections, as developed for
instance by \cite{ALMMT}. For isotropic models, this makes all states
$|\mu\rangle$ orthonormal.

Upon completion to a Hilbert space, a general state takes the form of
a countable superposition of $\exp(i\mu c)$ with
$\mu\in{\mathbb R}$:
\begin{equation} \label{States}
 \psi(c)= \sum_{I\in {\cal I}\subset {\mathbb R},\: {\rm countable}}
 f_I\exp(i\mu_Ic)
\end{equation}
such that $\sum_{I\in {\cal I}}|f_I|^2$ exists. These states form a
Hilbert space equivalent to the $\ell^2$-space formed by the
normalizable sequences $(f_I)_{I\in {\cal I}}$ for all countable
${\cal I}\subset{\mathbb R}$. In this form, isotropic states directly
result from the reduction of a full state, where all the holonomies
reduce to exponentials of $c$ with different exponents. Alternatively,
the Hilbert space may be characterized as the space of square
integrable functions $\psi(c)$ on the Bohr compactification
of the real line, a compact space
containing the real line densely and equipped with the measure
\[
 \int{\rm d}\nu (c)\psi(c)= \lim_{C\to\infty}\frac{1}{2C}\int_{-C}^C
 \psi(c){\rm d}c
\]
using the normal Lebesgue measure on the right-hand side. This Hilbert
space is non-separable.

Basic operators act by multiplication or differentiation:
\begin{eqnarray}
 \widehat{\exp(i\delta c)}|\mu\rangle&=& |\mu+\delta\rangle\\
\hat{p}|\mu\rangle&=& \frac{8\pi}{3}\gamma\ell_{\rm P}^2 \mu
|\mu\rangle
\end{eqnarray}
as it indeed follows from the unique holonomy-flux algebra in
the full theory as an induced representation. A Wheeler--DeWitt
representation, by contrast, would not be related to such a
representation of the full theory. 

Properties of the loop representation are markedly different from the
Wheeler--DeWitt one, but they closely mimick properties of the full
holonomy-flux representation:
\begin{itemize}
\item There is a discrete spectrum of $\hat{p}$. While there is a
  continuous range for $\mu$, all eigenstates $|\mu\rangle$ are
  normalizable. For a non-separable Hilbert space as we are dealing
  with here, normalizability of eigenstates does not imply that the
  eigenvalues form a countable subset of the real line. In such a
  situation, the normalizability condition is more general, for it is
  insensitive to what topology one uses on the set of
  eigenvalues. Even if any real number can appear as an eigenvalue,
  this set would be discrete if one uses a discrete topology of
  the real line (for instance, one where every subset is an open
  neighborhood). Normalizability of eigenstates will also be one of the
  crucial properties for consequences of flux spectra in loop quantum
  cosmology.
\item There is no operator for $c$, and only holonomies are
  represented. Trying to derive an operator for $c$ from holonomies,
  for instance by taking a derivative of the action of
  $\widehat{\exp(i\delta c)}$ by $\delta$ at $\delta=0$, fails because
  holonomies are not represented continuously in $\delta$:
  $\langle\mu|\widehat{\exp(i\delta c)}|\mu\rangle=
  \langle\mu|\mu+\delta\rangle=\delta_{\delta,0}$ is not continuous.
\end{itemize}
\cite{Bohr} and \cite{BohrWigner} discuss these and other properties
of the Bohr compactification as used in loop quantum cosmology, and
\cite{BohrADM} do so in a quantum cosmology based on ADM variables.

Since these are the same basic properties as they are realized for the
full holonomy-flux algebra, they have the same qualitative
implications for the dynamics. Once used for the construction of
operators such as the Hamiltonian constraint, they lead to specific
quantum-geometry corrections in loop quantum cosmology as they do in
the full theory. Specifically,
\begin{enumerate}
\item Only almost-periodic functions of $c$ are represented as
  operators on states (\ref{States}), and must be expressible as
  $\psi(c)= \sum_{I\in {\cal I}\subset {\mathbb R}, {\rm countable}}
  f_I\exp(i\delta_Ic)$.  The Hilbert space does not allow an action of
  $c$ on its dense subset of triad eigenstates; any appearance of $c$
  not of almost-periodic form, such as the polynomial in the isotropic
  Hamiltonian constraint, must be expressed in terms of
  almost-periodic functions by adding suitable higher-order
  corrections in $c$. They become significant for large values of the
  curvature, via $\ell_0\tilde{c}=c/{\cal N}^{1/3}$ for a regular
  distribution of edges where $\delta_I\sim {\cal N}^{-1/3}$.
\item The isotropic flux operator $\hat{p}$ has a discrete spectrum
  containing zero and so lacks a direct, densely defined inverse.
  Well-defined versions can be obtained via identities such as
\begin{equation}\label{InvTriadIso}
 \frac{i}{\delta}e^{i\delta c}\{e^{-i\delta c},|p|^{r/2}\}=
 \{c,|p|^{r/2}\}= \frac{4\pi\gamma Gr}{3} |p|^{r/2-1}\; {\rm sgn}p
\end{equation}
which mimick the crucial one (\ref{InvTriad}) used in the full theory.
For $0<r<2$ we are expressing an inverse of $p$ on the right-hand
side, but do not need an inverse on the left-hand side; well-defined
and even bounded operators for inverse powers of $p$ result, as
derived by \cite{InvScale}. Within this range, $r$ is unrestricted by
general considerations; it thus appears as an ambiguity parameter (see
\cite{Ambig} for further discussions of ambiguities).  Also here,
corrections to classical expressions arise, in this case for small
flux values $\ell_0^2 \tilde{p}=p/{\cal N}^{2/3}$ near the Planck
scale. From a quantization of the left-hand side of
(\ref{InvTriadIso}), eigenvalues can readily be derived. They have the
form $\frac{4}{3}\pi\gamma Gr |p|^{r/2-1}\alpha_r(p)$ where
$\alpha_r(p)\sim 1$ asymptotically for large $p/{\cal
  N}^{2/3}\gg\ell_{\rm P}^2$ while $\alpha_r(p)\to0$ rapidly for
$p\to0$, cutting off the divergence of $|p|^{r/2-1}$. The correct
semiclassical limit is guaranteed by $\alpha_r$ approaching one, which
it does from above irrespective of quantization ambiguities.
\end{enumerate}

\subsubsection{Difference equation}

The main quantity for which these considerations play a role is the
Hamiltonian constraint, in isotropic variables providing the
Friedmann equation
\[
 \frac{c^2\sqrt{|p|}}{\gamma^2}= \frac{8\pi G}{3}H_{\rm matter}
\]
with the matter Hamiltonian $H_{\rm matter}= V_0a^3\rho$.
Classically, its dependence on $c$ is via $c^2$, which is not almost
periodic. There are many ways to express $c^2$ in terms of
almost-periodic functions such that they approximate $c^2$ for small
curvature, $c\ll1$.  In general terms, properties of the quantum
representation space require the replacement of $c^2$ to be a bounded
function of $c$, given by a normalizable superposition
$\exp(\delta_Ic)$ with $\delta_I$ in a countable subset of the real
line. Each exponential acts by a shift of triad eigenvalues, rather
than a derivative operator, and so, following \cite{cosmoIV,IsoCosmo},
upon loop quantization the Hamiltonian constraint equation becomes a
difference equation for components of a wave function in the triad
representation.

A choice usually made is $c^2\sim \delta^{-2}\sin^2(\delta c)$,
but others are possible, constituting further ambiguities. A
Hamiltonian constraint operator
\[
 \hat{C}_{\rm iso}= \frac{3}{8\pi G\gamma^2\delta^2}\widehat{\sin
   \delta c}^2 \widehat{\sqrt{|p|}}- \hat{H}_{\rm matter}
\]
results. Expanding a state
$|\psi\rangle=\sum_{\mu}\psi_{\mu}|\mu\rangle$ in the triad eigenbasis
$\{|\mu\rangle\}_{\mu\in{\mathbb R}}$, $\hat{C}_{\rm iso}|\psi\rangle=0$
is equivalent to the difference equation
\begin{equation} \label{Diff}
  C(\mu+2\delta) \psi_{\mu+2\delta}-2 C(\mu)\psi_{\mu}+
  C(\mu-2\delta) \psi_{\mu-2\delta}= \frac{8\pi G}{3} 
\gamma^2\delta^2 \hat{H}_{\rm
    matter}(\mu)\psi_{\mu}
\end{equation}
as derived by \cite{IsoCosmo}, where $C(\mu)$ are eigenvalues of
$\widehat{\sqrt{|p|}}$, e.g.\ proportional to $\sqrt{|\mu|}$.
Additional matter fields have been suppressed in the notation, which
would provide further independent variables in $\psi_{\mu}$ acted on
by the matter Hamiltonian $\hat{H}_{\rm matter}(\mu)$. (Unless there
are curvature couplings, matter Hamiltonians depend on the densitized
triad but not on the connection; the right-hand side of (\ref{Diff}) is
then not a difference expression. Fermions, which require a coupling
to the gravitational connection, have been discussed in this context
by \cite{LRSParity}, and non-minimally coupled scalars by \cite{NonMin}.)

Due to the inequivalence of representations, this difference equation
replaces the differential Wheeler--DeWitt equation. At this dynamical
level, after choosing the representation of $c^2$ by almost-periodic
functions, the dynamics can be restricted to a separable subsector of
the initial non-separable Hilbert space. From this perspective, one
could have started directly from a separable Hilbert space by choosing
a specific countable set $\{\mu_I\}\subset{\mathbb R}$, such as
$\mu_I=I\mu_0$ with integer $I$ as originally defined by
\cite{IsoCosmo}. Many dynamical properties would follow in the same
way since non-separable features appear only at the kinematical
level. 

However, there is a good reason for dealing with a non-separable
kinematical Hilbert space, allowing for all real values of $\mu$: To
capture the most general and viable dynamics, one expects lattice
refinement to happen, with a discreteness scale changing dynamically.
Indeed, full Hamiltonian constraint operators create new edges and
vertices, changing the graph on which a state is defined, as
illustrated in Fig.~\ref{f:Local}. In an isotropic context, a trace of
this feature must still be left since the dynamical parameter $\delta$
came from the edge length $\ell_0$ used in elementary holonomies.
While $V_0$ in $\delta=\ell_0/V_0^{1/3}$ remains constant as a
mere auxiliary parameter introduced in the setup, $\ell_0$ must be
adapted to the geometry. Geometrical
distances, after all, increase as the universe expands and the region
of coordinate size $V_0$ has a growing geometrical size $V_0a^3$. If a
lattice state for a larger region has more sites, as suggested by the
creation of vertices, $\ell_0$ must shrink as $a$ grows; otherwise the
discrete size $a\ell_0$ would be blown up macroscopic. Here, the
evolutionary aspects capture expected properties of full physical
states, solving a vertex-creating Hamiltonian constraint, arranged by
volume eigenvalues as internal time. (From a more general perspective,
dynamical lattices and their role in quantum gravity have been
discussed by \cite{Weiss,UnruhTime,River}.) Then, also $\delta(\mu)$
must not be constant but rather a decreasing function. (Just as in
(\ref{F}), for the regularization of the isotropic $F_{ab}^i$, or the
$c^2$ in the Friedmann equation, by $\delta^{-2}\sin^2(\delta c)$, we
use the coordinate area $\ell_0^2$ of loops, not the geometrical area
$\ell_0^2a^2$ which might well be constant as $a$ changes.)

In summary, lattice refinement means that $\delta$ depends on the size
of the universe, and $\delta(\mu)$ is not a constant. The resulting
difference equations, as most generally formulated for anisotropic
models by \cite{SchwarzN}, are not equidistant and do not allow simple
restrictions to separable sectors.  A large class of different
dynamics can thus be formulated in each single model.\footnote{Several
improvised models have been formulated and sometimes analyzed in
detail, based on further ad-hoc assumptions to constrain
ambiguities. Since such an approach cannot capture the generic
behavior, specific details of the results may not be reliable.}
Consistency conditions do exist restricting the freedom even in
homogeneous models, but systematic investigations have only just
begun, as e.g.\ by \cite{Consistent}. While specific details of the
difference equations of loop quantum cosmology are not yet fully
determined, owing to quantization ambiguities, there are several key
generic features implying further properties. They are brought out
most clearly in sovable models.

\subsubsection{Harmonic cosmology}

Sometimes, solutions to the difference equations of loop quantum
cosmology, which are linear but with non-constant coefficients, can be
found. They may be studied numerically, and lead to interesting
issues especially in cases where they are not equidistant as
investigated so far by \cite{RefinedNumeric,RefinementNumeric}. But in
all cases, computing observables from the resulting wave functions and
arriving at sufficiently generic conclusions is challenging.

An effective treatment at this stage becomes much more powerful.
Initially, one may expect technical difficulties due to the non-linear
and non-polynomial nature of the Hamiltonian constraint obtained after
the loop replacement of $c^2$. Fortunately, isotropic loop quantum
cosmology offers an exactly solvable system where, in a specific
factor ordering of the constraint operator, the dynamics is
essentially free.  From the point of view of quantum dynamics, this
model is closely related to the harmonic oscillator of quantum
mechanics, which forms the basis of most of the perturbation-theory
framework for interacting quantum field theories. Similarly, the
harmonic model of loop quantum cosmology plays a crucial role for
perturbations in quantum gravity.

Specifically, according to \cite{BouncePert,BounceCohStates}, we
obtain the solvable model for spatially flat isotropic loop quantum
cosmology with a free, massless scalar, whose loop-modified
Hamiltonian constraint is
\[
\frac{4\pi G}{3}\frac{p_{\phi}^2}{p^{3/2}}= \sqrt{p}\gamma^{-2}
  \delta(p)^{-2} \sin^2(\delta(p)c) \sim \sqrt{p}c^2 +O(c^4)\,.
\]
Here, $\delta(p)$ (or ${\cal N}(p)=\delta(p)^{-3}$) is allowed to
depend on the triad, incorporating different refinement schemes of the
discrete structure.  For power laws $\delta(p)\propto p^x$, we
introduce $V=p^{1-x}/(1-x)$ and $J=p^{1-x}\exp(ip^{x}c)$ as
non-canonical basic variables forming an ${\rm sl}(2,{\mathbb R})$
algebra
\[
 [\hat{V},\hat{J}]=\hbar\hat{J}\quad,\quad{}
 [\hat{V},\hat{J}^{\dagger}]=-\hbar\hat{J}^{\dagger}\quad,\quad{}
 [\hat{J},\hat{J}^{\dagger}]=-2\hbar\hat{V}{}
\]
upon quantization.  Thanks to the free scalar, the dynamics is
controlled by the deparameterized Hamiltonian in $\phi$,
$\hat{p}_{\phi}\propto
|\frac{1}{2i}(\hat{J}-\hat{J}^{\dagger})|=:\hat{H}$.

This Hamiltonian is linear for all $x$, which for our linear algebra
of non-canonical basic variables implies solvability. (Again, the
absolute value is irrelevant for effective equations.) Following our
general analysis of effective descriptions, we can directly jump to
equations of motion for expectation values of the basic variables.
They are coupled to each other, but not to quantum variables such as
fluctuations as it normally occurs in interacting quantum systems.
Instead of a non-linear set of infinitely many differential equations,
we have a finitely coupled set of linear equations.  For the basic
variables, we have
 \begin{eqnarray*}
  \frac{{\rm d}\langle\hat{V}\rangle}{{\rm d}\phi}&=& 
\frac{\langle[\hat{V},\hat{H}]\rangle}{i\hbar}=
  -{\textstyle\frac{1}{2}}(\langle\hat{J}\rangle+ 
  \langle\hat{J}^{\dagger}\rangle)\\
  \frac{{\rm d}\langle\hat{J}\rangle}{{\rm d}\phi}&=& 
\frac{\langle[\hat{J},\hat{H}]\rangle}{i\hbar}=
  -{\textstyle\frac{1}{2}}\langle\hat{V}\rangle=
  \frac{{\rm d}\langle\hat{J}^{\dagger}\rangle}{{\rm d}\phi}\,.
\end{eqnarray*}

As a difference to the previous discussion of effective systems, we
now have to take into account the complex nature of our variable $J$.
Reality conditions are required for physical results, which would be
implemented by the physical inner product in a Hilbert-space
representation. Here, we are not using states but directly work with
expectation values and moments. The adjointness relation
$\hat{J}\hat{J}^{\dagger}=\hat{V}^2$ for our basic operators implies,
upon taking an expectation value, a reality condition relating
$|\langle\hat{J}\rangle|^2-\langle\hat{V}\rangle^2$ to moments of
$\hat{V}$ and $\hat{J}$. It turns out that the specific combination of
moments involved is constant in $\phi$-evolution, and that it is of
the order $\hbar$ for a semiclassical state. (For more details, see
\cite{BounceCohStates,BounceSqueezed}.) Just requiring that the state
is semiclassical only once, for instance at large volume, ensures that
the reality condition reads
$|\langle\hat{J}\rangle|^2-\langle\hat{V}\rangle^2=O(\hbar)$ at all
times. With this condition, the general solution is
\[
 \langle\hat{V}\rangle(\phi)=\langle\hat{H}\rangle\cosh(\phi-\beta)\quad,\quad
\langle\hat{J}\rangle(\phi)=-\langle\hat{H}\rangle(\sinh(\phi-\beta)-i)
\]
with the conserved $\langle\hat{H}\rangle$ and another integration
constant $\beta$.  For a large class of states we thus have an exact
realization of a bounce, with the volume bounded away from zero.
(Effective equations thus easily confirm the numerical
results of \cite{APS}, at least as far as physical expectation values
are concerned. For the dynamical behavior of moments, the generic
treatment of \cite{BeforeBB,BounceSqueezed} based on effective
equations shows crucial differences to the numerics done for only a
specific class of states.)

However, the system is harmonic, and so its behavior is not easily
generalizable to realistic models. While it is interesting that bounce
models can be derived in this way, taking this at face value without
realizing the solvable nature of the model would provide a view with
severe limitations. After all, the harmonic oscillator does not
provide general insights into quantum dynamics, and free quantum field
theories allow no glimpse at the rich features of interacting ones.
Only a systematic analysis of perturbations around the free model,
starting with interaction terms and then introducing inhomogeneities,
can provide a clear picture.

\subsubsection{Quantum Friedmann equation}

As already briefly discussed in the context of Wheeler--DeWitt quantum
cosmology, effective equations of quantum cosmology are necessarily of
higher dimension than the classical equations. Quantum degrees of
freedom such as fluctuations couple to expectation values in
non-solvable models, and no adiabatic or other regime has been
determined where quantum variables could completely be expressed in
terms of effective contributions depending only on the expectation
values. Quantum variables are true degrees of freedom, of significance
for the dynamics. Without a clear vacuum state to expand around,
moreover, suitable states are less restricted than they are for
low-energy effective actions. Via initial values for moment equations,
this leaves the state dependence as a crucial contribution of
equations, to be taken into account for sufficiently general
conclusions.  Since not much is known about details of the quantum
state of the universe, results should be sufficiently insensitive to
its properties.

Perturbation theory by the moments, including them order by order, is
required for any model which is not harmonic. Examples are models with
non-vanishing spatial curvature or a cosmological constant as analyzed
by \cite{Recollapse} (or numerically by \cite{NegCosNum}). Another
class of important models is that where the scalar is no longer free
or at least acquires a mass term. We are then dealing with a
time-dependent Hamiltonian in the $\phi$-evolution, and $\phi$ may not
even provide a good internal time if the potential leads to turning
points of $\phi(\tau)$. At this stage, a more general analysis is
required which is now available in the form of effective constraints
as per \cite{EffCons} and \cite{EffConsRel}: we do not have to
deparameterize the system before quantizing it or computing effective
equations. We can directly compute effective constraints and analyze
them to find the physical quantum phase space.  As one of the results,
we can apply the deparameterized framework even if there is a
potential provided it changes sufficiently slowly. This is exactly the
case of interest for inflationary or other early-universe cosmology,
and so we can derive effective equations for these situations.

In the presence of a potential, the Friedmann equation receives the
following quantum corrections:
\[
 \left(\frac{\dot{a}}{a}\right)^2 = \frac{8\pi G}{3}\left(\rho
 \left(1-\frac{\rho_Q}{\rho_{\rm crit}}\right)
 \pm\frac{1}{2}\sqrt{1-\frac{\rho_Q}{\rho_{\rm crit}}} 
\eta (\rho-P)+ \frac{(\rho-P)^2}{\rho+P}\eta^2
\right)
\]
as derived by \cite{QuantumBounce,BounceSqueezed}. (The first term,
corresponding only to the solvable model with a free scalar, was
earlier obtained by \cite{SemiClassEmerge,RSLoopDual}. It sums up all
higher-order corrections due to holonomies, as explicitly expanded by
\cite{DiscCorr}. In the usual terminology of quantum field theory, it
represents the tree-level approximation since its main form is
independent of quantum back-reaction.) Here, $P$ is pressure, $\eta$
parameterizes quantum correlations and
\[
 \rho_Q:=\rho+\epsilon_0 \rho_{\rm crit}+ (\rho-P) \sum_{k=0}^{\infty}
\epsilon_{k+1}
 \left(\frac{\rho-P}{\rho+P}\right)^k
\]
is a quantum-corrected energy density with fluctuation parameters
$\epsilon_k$. The critical density $\rho_{\rm crit}=3/8\pi
G\gamma^2\delta^2V_0^{2/3}a^2$ results from the loop quantization,
bringing in the scale $\delta(a)$ whose form is determined by lattice
refinement.

This equation contains all quantum variables in $\eta$ and $\rho_Q$,
subject to their own dynamics. Understanding the behavior of
generic universe models in loop quantum cosmology requires a
high-dimensional dynamical system to be analyzed. Only a few simple
cases allow strict conclusions about the presence of a bounce. If
$\rho=P$ (a free, massless scalar) we recover the solvable scenario.
For $\rho+P$ large (large $p_{\phi}$, i.e.\ kinetic domination), all
corrections involving quantum variables are supressed; at least for a
certain amount of time, the behavior can be expected to be close to
the solvable one. 

In general, however, quantum variables may yank the universe away from
the simple bouncing behavior of the solvable model. While the
fundamental dynamics based on difference equations for states remains
non-singular, following \cite{Sing}, the general effective one is
still unclear. Volume expectation values could asymptotically approach
a non-zero constant without bouncing back to large values, a picture
which might resemble that of the emergent universe of
\cite{Emergent,Emergent2}. Most likely,
generic evolution would bring us to a highly quantum regime where
effective equations or simple near-solvable models break down.
Ultimately, the fate of a universe can be studied only based on the
fundamental difference equations, but effective equations can show
well how such severe states are approached.

\subsubsection{The interplay of different quantum corrections}

Loop quantum cosmology leads to different types of quantum corrections
in its effective equations: holonomy corrections, inverse triad
corrections, and quantum back-reaction. Depending on parameters, they
may play different roles in any given regime, and some of them might
dominate. What exactly happens can be determined only with a
sufficiently general parameterization of correction terms. Here, all
ambiguities and effects such as lattice refinement must be considered.

Especially holonomy and inverse triad corrections are often closely
related to each other: they both result from the discrete
geometry, although in different ways. Comparing them can thus provide
more restrictions on free parameters of the full theory than any one
of the corrections would allow individually. They are both related to
the size of elementary building blocks of a discrete geometry. For a
nearly isotropic distribution in a region of coordinate size $V_0$,
the classical volume on the left-hand side of
\[
 V_0a(\phi)^3= {\cal N}(\phi)v(\phi)
\]
is replaced by the right-hand side in the discrete picture. This
elementary relationship first tells us that, dynamically, there are
two free functions in the discrete picture for one free function
$a(\phi)$ in the continuum picture: the size $v(\phi)$ and number
${\cal N}(\phi)$ of discrete sites. Both of them typically change in
internal time: the lattice they describe is being refined as time goes
on.

As derived earlier, one of these parameters, $v$, enters the basic
expressions for holonomies
\[
 \exp(i\ell_0c/V_0^{1/3})=\exp(ic/{\cal
  N}^{1/3})=\exp(i\gamma v^{1/3}\dot{a}/a)
\]
using $c=V_0^{1/3}\tilde{c}= V_0^{1/3}\gamma\dot{a}$.  Inverse-triad
corrections in isotropic models provide a correction factor $\alpha$
depending on $p/{\cal N}^{2/3}= (V_0/{\cal N})^{2/3}a^2= v^{2/3}$. For
$v^{1/3}\sim \ell_{\rm P}$, inverse-triad corrections differ strongly
from the classically expected value $\alpha=1$ (while holonomies
remain nearly equal to the classically linear $\dot{a}/a$ for Hubble
distances much larger than $v^{1/3}$).  Here, quantum corrections can
often be constrained. Since two different corrections depend on one
parameter, their interplay can provide important synergistic effects
in ruling out possible values, or even entire corrections.\footnote{We
can also note that all quantum geometry corrections depend only on
$v=V_0a^3/{\cal N}$. As the size of discrete building blocks, it is
insensitive to changing $V_0$ (both $V_0$ and ${\cal N}$ change
proportionally to $V_0$) or coordinates ($V_0a^3$ and ${\cal N}$ are
scaling invariant). Just like the classical equations, quantum
corrections are invariant under these rescalings. For the scale
factor, inverse-triad corrections become significant at a
characteristic scale $a_*= ({\cal N}/V_0)^{1/3} \ell_{\rm P}$ related
to the Planck length.  While $a_*$ is not scaling independent, it
scales in the same way as $a$. Comparisons such as $a>a_*$ or $a<a_*$
to demarcate the classical and the strongly quantum regime are thus
meaningful.}

Phenomenologically, $v$ shows up in the critical density of the
effective Friedmann equation: $\rho_{\rm crit}=3/8\pi G \gamma^2
v^{2/3}$. At this value, we would find the bounce of a universe
sourced by a free, massless scalar, and even in other models its value
affects details of the dynamics. This density is often assumed
Planckian, which means that $v\sim \ell_{\rm P}^3$. (From black-hole
entropy calculations, as developed by \cite{ABCK:LoopEntro},
\cite{KaulMaj}, \cite{Gamma} and \cite{Gamma2}, $\gamma\sim 0.2$.) But
then, inverse triad corrections are large. If $v$ is constant, which
is also often assumed, inverse-triad corrections would be large at all
times, even at large total volume. A Planckian constant value of $v$
is thus clearly ruled out, not by holonomy corrections alone but by a
consistent combination with inverse-triad effects. Further
phenomenological analysis is in progress, by
\cite{RefinementInflation,RefinementMatter,SuperInflTensor,BounceTensor,TensorHalfII,TensorSlowRollInv,TensorSlowRoll,SuperInflPowerSpec}.
Much stronger consistency conditions can be expected when
inhomogeneities are included, to which we will turn next.

\section{Consistent dynamics}
\label{s:Cons}

A space-time covariant theory is a gauge theory with the gauge group
given by space-time diffeomorphisms. As a consequence, the local
conservation law $\nabla^{\mu}(G_{\mu\nu}-8\pi GT_{\mu\nu})$ holds for
the Einstein tensor $G_{\mu\nu}$ and the stress-energy tensor
$T_{\mu\nu}$ of matter. Not all components of Einstein's equation are
thus independent, and some can be derived from the others. For
consistency, effective equations of quantum gravity must show the same
form of dependencies, or allow the same number of local conservation
laws, or else degrees of freedom would be too much constrained
to have a chance of providing the correct classical limit.

Consistency is automatically satisfied if one has space-time covariant
higher-order corrections in Lagrangian form. At the Hamiltonian level,
however, covariance is more difficult to check due to the lack of a
direct action of space-time diffeomorphisms. There is rather a
splitting into spatial diffeomorphisms acting as usual, and an
independent generator deforming spatial constant-time slices within
space-time.  As in relativistic systems, all generators are
constraints, taking center stage in a canonical analysis: The
conservation law implies that the components $G_{0\nu}-8\pi GT_{0\nu}$
of Einstein's equation are only of first order in time derivatives since
$\partial_0(G^0_{\nu}-8\pi GT^0_{\nu})$ must equal terms of at most
second order in time. The same equation shows that these constraints
are automatically preserved in time, given that
$\partial_0(G^0_{\nu}-8\pi GT^0_{\nu})$ depends only on the Einstein
equations at fixed time and their spatial derivatives. If Einstein's
equation holds at one time, time derivatives of the components
$G_{0\nu}-8\pi GT_{0\nu}$ vanish; the constraints can consistently be
imposed at all times. Also this condition must be preserved for
effective equations.  Otherwise, the system is overdetermined with
more equations than unknowns and not enough consistent solutions would
result.

From this perspective, the covariance condition
takes the following form at the Hamiltonian level. For a generally
covariant system, evolution in coordinate time is generated by the
constraints, and the constraints must themselves be preserved in time.
Thus, the gauge transformation $\delta_{\epsilon^{\mu}} H[N^{\nu}]=
\{H[N^{\nu}],H[\epsilon^{\mu}]\}$ of a combination
$H[N^{\nu}]:=\int{\rm d}^3x N^{\nu}(G_{0\nu}-8\pi GT_{0\nu})$ must
vanish for all space-time vector fields $N^{\nu}$, $\epsilon^{\mu}$
when the constraints are satisfied. These Poisson-bracket relations
provide algebraic conditions for the constraints to be consistent,
forming a so-called first-class system. Specifically for general
relativity, we have
\begin{equation} \label{ConstrAlg}
 \{H[N^{\mu}],H[M^{\nu}]\}= H[K^{\mu}]
\end{equation}
with $K^0= {\cal L}_{M^{a}}N^0- {\cal L}_{N^a}M^0$ and $K^a= {\cal
  L}_{M^a}N^a- {\cal L}_{N^a}M^a +
q^{ab}(N^0\partial_bM^0-M^0\partial_b N^0)$.  It is a concise
expression for space-time covariance, and is insensitive to the
specific dynamics: the algebra is the same for all higher-curvature
actions (or with all kinds of matter), even though the specific
constraint functionals do change. (A recent Hamiltonian analysis of
higher-curvature actions has been performed by \cite{HigherCurvHam},
showing the constraint algebra explicitly.)

The canonical consistency requirement of an anomaly-free, covariant
theory then states that the algebra of effective constraints must
remain first class. As long as this algebraic condition is satisfied,
the theory is completely consistent.  The system of constraints must
remain first class after including quantum corrections, but the
specific algebra may change. Canonically, the realm of consistent
theories is larger than at the Lagrangian level: While corrections to
the action may be of higher-curvature form, they all give rise to
exactly the same constraint algebra. The Hamiltonian level, on the
other hand, allows changes in the algebra as long as it remains first
class; the theory may be consistently deformed. Thereby, changes to
the quantum space-time structure can be captured. Whether or not
non-trivial consistent deformations exist, and how their covariance
can be interpreted, is a matter of analysis in specific quantum
gravity models. We will illustrate the results of several models in
loop quantum cosmology in the following section.

\subsection{Cosmological perturbations}

The general issues of constrained systems in gravity can easily be
seen already for linear cosmological perturbations. For scalar modes,
perturbing the lapse function as $N(t)(1+\phi(x,t))$ and the scale
factor as $a(t)(1-\psi(x,t))$, and with a scalar matter source, the
linearized components of Einstein's equation, in conformal time and
longitudinal gauge, read
\begin{eqnarray}
\partial_c\left(\dot\psi+{\cal H}\phi\right)&=&4\pi G\dot{\bar\varphi}
\partial_c\delta\varphi \label{PertI}\\
\nabla^2\psi-3{\cal H}\!\left(\dot\psi+{\cal H}\phi\right)&=&4\pi
G\!\left(\!\dot{\bar\varphi}\delta\dot\varphi-\dot{\bar\varphi}^2\phi+a^2
V_{,\varphi}(\bar\varphi)\delta\varphi\!\right) \label{PertII}\\
\mbox{}\hspace{-3mm}\ddot\psi+{\cal H}\left(2\dot\psi+\dot\phi\right)
+\left(2\dot{\cal H}+{\cal H}^2\right)\phi&=&4\pi
G\left(\dot{\bar\varphi}\delta\dot\varphi-a^2
V_{,\varphi}(\bar\varphi)\delta\varphi\right)\label{PertIII}\\
\partial_c\partial^c(\phi-\psi)&=&0 \label{PertIV}
\end{eqnarray}
with the conformal Hubble parameter ${\cal H}$. This set of equations
is accompanied by the linearized Klein--Gordon equation for
$\varphi=\bar{\varphi}+\delta\varphi$. The first two lines are of
first order in time, and thus pose constraints for initial values.
They are the linearized diffeomorphism constraint (\ref{PertI}) and
the Hamiltonian constraint (\ref{PertII}). The next two lines are the
diagonal and off-diagonal components of the spatial part of Einstein's
equation, also linearized around Friedmann--Robertson--Walker. Using
the background equations for ${\cal H}$ and the background scalar
$\bar{\varphi}$, one can explicitly derive the Klein--Gordon equation
from the rest. The system is thus overdetermined, but in a consistent
way. Canonically, this follows from the fact that all equations result
from a set of first-class constraints.

In these equations, perturbations $\phi$, $\psi$ and $\delta\varphi$
are directly used for the metric and the scalar. These quantities are
subject to gauge transformations under changes of space-time
coordinates, which also change the gauge. So far, the equations as
written are in longitudinal gauge where the space-time metric remains
diagonal under perturbations. More generally, scalar modes can be
perturbed also in the off-diagonal part of the spatial metric by
$\delta q_{ab}= \partial_a\partial_bE$, or in the space-time part by a
perturbed shift vector $\delta N^a=\partial^aB$. Two new scalar
perturbations $E$ and $B$ are introduced in addition to $\phi$ and
$\psi$, and all four transform into each other under coordinate
changes. For linear coordinate changes, however, the combinations
\begin{equation}
\Psi=\psi-{\cal H}(B-\dot E)
\end{equation}
and
\begin{equation}
\Phi=\phi+\left(B-\dot E\right)^{\scriptscriptstyle\bullet}+{\cal H}(B-\dot E)
\end{equation}
and a similar one for the matter perturbation remain unchanged as
specified by \cite{Bardeen}; they form gauge-invariant observables of
the linear theory. From a canonical perspective, these combinations
are invariant under the flow generated by the constraints.

The ungauged evolution equations, containing $E$ and $B$ in addition
to $\phi$ and $\psi$, can be expressed completely in terms of the
gauge-invariant variables, as required on physical grounds.  Here, the
first-class nature of constraints is important which ensures that the
constraint equations, and thus (\ref{PertI}) and (\ref{PertII}), are
gauge invariant.  Consistency as well as gauge invariance of the
equations of motion is thus guaranteed by having a first-class algebra
of constraints. Also here, the first-class nature serves as a complete
requirement for covariance.

\subsection{Gauges and frames}

A first-class algebra of constraints ensures that physical evolution
can be formulated fully in gauge-invariant terms. By the same property
of the constrained system, also frame-independence is guaranteed: The
gauge algebra corresponds to deformations of the spatial hypersurfaces
given by a time function $t={\rm const}$. Gauge invariant variables
are insensitive to these deformations, and thus to the choice of time.
But a time function (without specifying which spatial coordinates are
to be held fixed) does not uniquely tell us how to take time
derivatives for the equations of motion; this is only accomplished
when we also specify a time-evolution vector field $t^a$ (such that
$t^a\partial_at=1$). For a given time function, there are many choices
for $t^a$, and the time-evolution vector field may be changed
independently of the foliation.

A fixed foliation of space-time into spatial slizes with unit normals
$n^a$ allows us to express the freedom in choosing time-evolution
vector fields by the lapse function $N$ and the shift vector $N^a$
(with $N^an_a=0$) such that $t^a=Nn^a+N^a$. These are exactly the
functions which appear in the constraints generating evolution: for a
fixed choice of $t^a$, or $N$ and $N^a$, Hamiltonian equations of
motion are $\dot{f}= {\cal L}_{t^a}f=\{f,H[N,N^a]\}$ for any
phase-space function $f$.  Since $N$ and $N^a$ appear as multipliers
of first-class constraints, they can be chosen arbitrarily (except
that we would like $N>0$ for evolution toward the future). For a
consistent set of first-class constraints, we can thus choose the
frame freely, and different frame choices are guaranteed to produce
consistent results. (In a space-time treatment, observables which are
gauge as well as frame-independent can be derived following
\cite{EB,BDE}. In a reduced phase-space treatment, where one uses
observables solving the classical constraints, one is working at a
gauge-invariant, but not automatically frame-independent level.)

In classical relativity, cosmological perturbation equations can often
be derived in much simper ways when a space-time gauge is chosen, such
as the longitudinal gauge above or the uniform one where only matter
fields are perturbed. Since gauge transformations are known to
correspond to space-time coordinate transformations, one can directly
verify that such gauges are possible. Moreover, choosing a gauge
before deriving equations of motion from an action or Hamiltonian is
equivalent to choosing a gauge in the general equations of motion. The
situation is thus completely unambiguous.

When quantum effects are included, either in a full quantum theory or
in an effective manner, the constraints change by quantum correction
terms. Equations of motion change, as
expected, and so do the form of gauge-invariant expressions since it
is the constraints which generate gauge transformations. In such a
situation, gauge transformations and suitable gauge fixings can be
analyzed only after the quantization has been performed and the
corrected constraints are known. If gauge fixings are employed before
quantization or before determining effective constraints, the choice
of gauge fixing may not be compatible with the resulting corrected
gauge transformations. Moreover, choosing different gauge fixings
before doing the same kind of quantization would in general lead to
different final results, making the procedure ambiguous even beyond
unavoidable quantization ambiguities. Similarly, a reduced phase space
quantization is based on frame-fixing, although no gauge need be
fixed.

For the different approaches used in relation to loop quantum gravity,
several examples exist. \cite{UniformDisc} develop methods to deal
with a discretization possibly breaking gauge symmetries. Similar
methods have then be used by \cite{SphSymmUniform} in spherically
symmetric models with (partial) gauge fixing. Also the hybrid
quantizations of Gowdy models by \cite{Hybrid} rely on gauge fixing of
the inhomogeneous generators. \cite{PerfectAction} construct
discretized theories in three dimensions, respecting the space-time
gauge, but argue in \cite{BrokenAction} that this may not be possible
in four space-time dimensions. \cite{PFTLoop} and \cite{PPFT} quantize
2-dimensional parameterized models of field theories by the Dirac
procedure and represent observables on the resulting physical Hilbert
space. In this treatment, the discretization does not lead to
inconsistencies but possibly to deformations of classical algebras of
observables.  Finally, reduced phase space methods fixing the frame by
referring to an extra dust field are developed by
\cite{BKdustI,BKdustII} for cosmological perturbations, and by
\cite{BKdustLTB} in spherical symmetry. Other treatments for
cosmological perturbations are used, e.g., by \cite{HolonomyInfl} and
\cite{BounceCMB} in gauge-fixed versions and by \cite{PuchtaMaster} in
a frame-fixed (reduced phase-space) way. \cite{LQCStepping} provide a
proposal by which cosmological perturbations, taking into account
space-time discreteness, might be implementable by a consistent
first-class algebra of constraints. So far, this has been realized for
coupling two independent homogeneous patches.

The only valid treatment of a complicated gauge theory is by working
without restrictions of the gauge throughout the quantization
procedure, until the final gauge algebra resulting from the corrected
constraints has been confirmed to be consistent. Here, the anomaly
problem, confirming that a consistent deformation is realized, must be
faced head-on and cannot be evaded. In the final equations one may
choose one of the allowed gauges for further analysis, but gauges
cannot be used to simplify the quantization. In the rest of this
exposition, we follow these lines to illustrate the consistency of
several effective sets of constraints incorporating some of the
discreteness effects of loop quantum gravity.

\section{Consistent effective discrete dynamics}

If loop quantum gravity has a chance of being a viable quantum theory
of gravity, the form of discrete quantum geometry it implies must give
rise to effective dynamical equations satisfying the consistency
conditions of a covariant theory. At the Hamiltonian level, this
requires a first-class algebra of constraints. After the preparation
in the preceding sections, we can now see what specific models
indicate.

\subsection{Constraint algebra}

Consistent deformations implementing the effects
of loop quantum gravity have been found in several different cases.
Most of them use inverse-triad corrections,
which have been incorporated successfully in spherically symmetric
models by \cite{SphSymmPSM,LTBII} as well as linear perturbations
around spatially flat Friedmann--Robertson--Walker models by
\cite{ConstraintAlgebra,ScalarGaugeInv}. The situation for holonomy
corrections is more restrictive; here, certain versions have been
realized in spherically symmetric models by \cite{JR} as well as for
linear tensor and vector modes around Friedmann--Robertson--Walker
models in \cite{vector,tensor}.  However, so far no inhomogeneous
model has been found where holonomy corrections in a complete form
would be consistent. Here, the requirement of anomaly-freedom appears
very restrictive.

Inverse-triad corrections have been implemented consistently in
several settings, and so are not ruled out by consistency
considerations. In particular, corrections from the discreteness of
quantum geometry are allowed and do not necessarily spoil covariance.
The specific form of their implementation then tells us if and how
space-time structures have to change due to quantum effects.

Based on formulas such as (\ref{InvTriad}), inverse-triad corrections
arise for any term in the Hamiltonian constraint bearing components of
the inverse densitized triad, such as $1/\sqrt{|\det E^a_i|}$.
\cite{ConstraintAlgebra} have consistently implemented these
corrections for linear inhomogeneities, where the corrected constraint
algebra is of the form (\ref{ConstrAlg}) but with
\[
 K^a={\cal L}_{M^a} N^a- {\cal L}_{N^a}M^a+
\bar{\alpha}^2\bar{N}a^{-1/2}
 \partial^a (\delta M^0-\delta N^0)
\]
while $K^0={\cal L}_{M^a}\delta N^0- {\cal L}_{N^a}\delta M^0$ retains
its classical form.  In addition to the contribution $\bar{N}a^{-1/2}
\partial^a (\delta M^0-\delta N^0)$, which is expected classically for
a linearization around Friedmann--Robertson--Walker models with
$N_i=\bar{N}+\delta N_i$, there is the function $\bar{\alpha}$
(depending on the background scale factor $a$) arising from
inverse-triad corrections. An algebra of the same form arises for
spherically symmetric models, with different matter couplings as
discussed by \cite{JR}. The constraint algebra is anomaly-free: the
system of constraints remains first class. But it is not exactly the
classical algebra, and thus deformed. Inverse-triad corrections from
loop quantum gravity cannot amount to higher-curvature corrections to
the action since this would leave the constraint algebra unchanged.
Rather, these corrections can only be understood as deforming local
space-time symmetries while keeping covariance realized.

\subsection{Cosmological perturbations}

Consistent versions for quantum-corrected constraints allow one to
analyze their implications for the dynamics. When constraints are
corrected, not just evolution equations change but also the gauge
transformations generated by the constraints. Thus, expressions for
gauge-invariant observables depend differently on perturbations of the
fields, which by itself may give rise to new effects. Other
implications then follow from studying the dynamical evolution of
gauge-invariant observables. (Some quantum-gravity corrections have
been implemented in gauge-fixed versions. They are formally
consistent, but quite arbitrary in their implementation. For instance,
it remains unclear how different gauge-fixings, all done before
quantization, might affect the results. Moreover, some physical
effects due to corrections to gauge-invariant observables can easily
be overlooked.)

For linear perturbation equations around Friedmann--Robertson--Walker
models, cosmological perturbation equations are the main application
of consistent deformations. With a consistent deformation,
perturbation equations form a closed set and can be written fully in
terms of gauge-invariant variables. For
inverse-triad corrections, as derived by \cite{ScalarGaugeInv}, they
take the form
\[
\partial_c\left(\dot\Psi+{\cal H}(1+f)\Phi\right)=\pi
G\frac{\bar{\alpha}}{\bar{\nu}}\dot{\bar{\varphi}} \partial_c\delta\varphi^{\rm GI}
\]
as the corrected time-space part of Einstein's equation,
\begin{eqnarray*}
&&\Delta(\bar{\alpha}^2
\Psi)-3{\cal H}(1+f)
\left(\dot\Psi+{\cal H}\Phi(1+f)\right)\\
&=&4\pi
G\frac{\bar{\alpha}}{\bar{\nu}}(1+f_3)
\left(\dot{\bar{\varphi}}\delta\dot\varphi^{\rm
GI}-\dot{\bar{\varphi}}^2(1+f_1)\Phi
+\bar{\nu} a^2 V_{,\varphi}(\bar{\varphi})
\delta\varphi^{\rm GI}\right)
\end{eqnarray*}
as the corrected time-time part, and
\begin{eqnarray*}
&&\ddot\Psi+{\cal H}\left(2\dot\Psi\left(1-\frac{a}{2\bar{\alpha}}
\frac{{\rm d}\bar{\alpha}}{{\rm d}a}\right)+\dot\Phi(1+f)\right)
+\left(2\dot{\cal H}+{\cal H}^2\left(1+
\frac{a}{2}\frac{{\rm d}f}{{\rm d}a} -
\frac{a}{2\bar{\alpha}}\frac{{\rm d}\bar{\alpha}}{{\rm d}a}\right)\right)
\Phi(1+f)\\
&=&4\pi G\frac{\bar{\alpha}}{\bar{\nu}}
\left(\dot{\bar{\varphi}}\delta\dot\varphi^{\rm
GI}-a^2\bar{\nu} V_{,\varphi}(\bar{\varphi})\delta\varphi^{\rm GI}\right)
\end{eqnarray*}
as the diagonal space-space part. All corrections $f$, $f_1$, $f_3$
and $h$ below are determined from the basic inverse-triad corrections
$\bar{\alpha}$ (for the gravitational part of the constraint) and
$\bar{\nu}$ (for the kinetic term of the matter part). Since these are
background corrections, their form can easily be computed in isotropic
models, suitably parameterized for all ambiguities and lattice
refinement.

The off-diagonal space-space part also implies a non-trivial equation,
with an unexpected consequence: While the classical analog would
simply identify $\Phi=\Psi$, the corrected equation implies
$\Phi=\Psi(1+h)$ with a quantum correction by $h\not=0$.
This may be interpreted as an
effective anisotropic stress contribution, but it results from a
correction to quantum gravity, not to matter.

As a second implication, we may have non-conservation of power
on large scales as pointed out by
\cite{InhomEvolve,ScalarGaugeInv}.\footnote{{\em Note added:} Recently
it turned out that power, due to an unexpected cancellation, is
conserved on large scales for the currently available equations with
inverse-triad corrections, as derived by \cite{InflObs}. So far,
however, there is no general argument for the conservation of power in
the presence of modified space-time structures. The analysis of
\cite{InflConsist} has shown that interesting effects for potential
observations mainly arise from corrections to the running of indices.}
This may be important for inflationary structure formation, where the
long evolution times while modes are outside the Hubble radius would
make even a weakly changing size of the overall power
significant. Both of these effects are difficult to see in gauge-fixed
treatments, such as the longitudinal or uniform gauge, or in
frame-fixed versions based on reduced phase space quantizations.
Since scalar cosmological perturbations have been consistently
formulated only for inverse-triad corrections, no version is yet able
to evolve perturbations through a bounce for which holonomy
corrections are required. (Implementing holonomy corrections only for
the background leads to inconsistent evolution equations for
inhomogeneities.)

In addition to inverse-triad corrections, holonomies and quantum
back-reaction must be implemented to obtain a full picture from loop
quantum gravity. While consistent deformations are not yet known for
the latter two corrections, \cite{StructureGen} have formulated
quantum back-reaction in a cosmological setting. In general, in
quantum gravity this requires the inclusion of moments between all
degrees of freedom of gravity and matter, including quantum
correlations between them. By setting the quantum variables of gravity
as well as its quantum correlations to zero, one obtains the effective
equations of quantum field theory on a curved space-time as a limit.
Including leading order corrections from the gravitational quantum
variables provides quantum field theory on a quantum space-time. While
such limiting cases can be realized explicitly at the effective level,
much still remains to be done for a detailed analysis of specific
properties.

\subsection{Causality}

Several examples illustrate the importance of a consistent
constraint algebra, rather than just any deformation as 
allowed in gauge-fixed or frame-fixed treatments. We have already seen
that some effects in cosmological perturbation equations can be obtained
only when neither gauge nor frame are fixed before the theory is
quantized or effective constraints are derived.  Another example is
the realization of causality, as \cite{tensor} studied it by comparing
the propagation of gravitational waves to that of light.

With quantum-gravity corrections, the gravitational as well as the
Maxwell Hamiltonian change compared to the classical expressions,
affecting evolution equations and their plane-wave solutions. The
gravitational contribution to the Hamiltonian constraint for
inverse-triad corrections is
\[
 H_{\rm G} = \frac{1}{16\pi G} \int_{\Sigma} \mathrm{d}^3x \alpha(E^a_i)
\frac{ E^c_jE^d_k}{\sqrt{\left|\det E\right|}}
\left({\epsilon_i}^{jk}F_{cd}^i -2(1+\gamma^{2}) 
K_{[c}^j K_{d]}^k\right)
\]
implying the linearized wave equation
\[
\frac{1}{2}\left( \frac{1}{\alpha} \ddot{h}_a^i +
  2\frac{\dot{a}}{a} \left(1-\frac{2a{\rm d}\alpha/{\rm
        d}a}{\alpha}\right)\dot{h}_a^i -\alpha\nabla^2 h_a^i\right)
= 8\pi G \Pi_a^i
\]
for the tensor mode $h^i_a$ on a cosmological background with scale
factor $a$ and source-term $\Pi_a^i$.  By a plane-wave ansatz, we
derive the dispersion relation $\omega^2 = \alpha^2 k^2$ for
gravitational waves. Since $\alpha>1$ for perturbative corrections,
there is a danger of gravitational waves being super-luminal.

With these corrections, gravitational waves are faster than light on a
classical background. For a meaningful comparison, however, we should
use the speed of gravitational waves in relation to the physical speed
of light on the same background, which should receive quantum
corrections, too. Here, the Hamiltonian is
\[
H_{\rm EM} = \int_{\Sigma}\mathrm{d}^3x
\left(\alpha_{\rm EM}(q_{cd}) \frac{2\pi}{\sqrt{\det q}} E^aE^b q_{ab}
  + \beta_{\rm EM}(q_{cd})\frac{\sqrt{\det q}}{16\pi} F_{ab} F_{cd}
  q^{ac} q^{bd} \right)
\]
with two correction functions $\alpha_{\rm EM}$ and $\beta_{\rm EM}$
kinematically independent of the gravitational correction $\alpha$.
From the wave equation
\[
 \partial_t \left( \alpha_{\rm EM}^{-1} \partial_t A_a\right) -
  \beta_{\rm EM}\nabla^2 A_a = 0
\]
we obtain the dispersion relation $\omega^2 = \alpha_{\rm EM}
\beta_{\rm EM} k^2$, which also is ``super-luminal'' compared to
the classical speed of light.

Working out the requirements for anomaly-freedom with these two
contributions to the Hamiltonian constraint, as done by \cite{tensor},
we find $\alpha^2=\alpha_{\rm EM}\beta_{\rm EM}$
and the dispersion relations are equal.  Physically, for comparisons
of speeds on the same background, there is no super-luminal
propagation. In a gauge-fixed or frame-fixed treatment, by comparison,
one could have chosen the correction functions independently of
another since tensor modes of the gravitational field and the electric
field make up independent physical observables. A gauge- or frame-fixed
treatment could easily produce corrected equations violating
causality, but this is ruled out by a complete treatment.

\section{Outlook: Future dynamics}

To probe a quantum theory of gravity or even arrive at predictions one
must evaluate its dynamics in detail. For low-energy effects, leading
corrections to classical equations must be derived. The best tool for
systematic investigations in such cases is that of effective
descriptions, providing the evolution of expectation values of
observables in a physical state. At the same time, they can tell us
much of the entire behavior of physical states.

Hamiltonian effective descriptions can be applied directly to
canonical quantum gravity and cosmological models. In particular,
typical implications such as the discreteness of spatial or space-time
structures can then be probed, or first ensured to be consistent at
all. Several examples in loop quantum cosmology have demonstrated that
discreteness corrections can indeed be implemented consistently,
leaving the theory covariant but deforming its local space-time
symmetries. Future work must ensure that this is indeed possible for
the full theory of loop quantum gravity and its effective constraints.

While isotropic solvable models of loop quantum cosmology suggest a
role of bouncing cosmologies for potential scenarios, no consistent
set of equations to evolve inhomogeneities through a bounce has been
found. The only available options so far make use of gauge (or
frame) fixings before quantization, and thus miss crucial aspects of
space-time structures. Any mismatch of growing modes in the
collapse and expansion phases can easily be enhanced by cosmic
evolution, providing opportunities for potential observations but also
requiring extreme care in finding fully consistent
equations. Inhomogeneous cosmological scenarios remain uncertain, and
with it follow-up issues such as the entropy problem.

What models investigated so far suggest, in many different versions,
is that the classical algebra of space-time diffeomorphisms is
deformed but not violated. It is then clear that quantum corrections
cannot merely amount to higher-curvature terms in an effective action,
although such terms may appear, too. Instead,
quantum structures of space-time must change by quantum effects. While
space-time covariance is no longer realized in the
standard sense, from the Hamiltonian perspective the effective
theories remain completely consistent and covariant with an underlying
first-class algebra of gauge generators.  As indicated by the initial
quote from \cite{DiracHamGR}, this sense of covariance, from a general
perspective, is the appropriate one. Its consistent implementation,
without fixing gauge or frame, can tell us a great deal about the
fundamental structures of space and time.

\section*{Acknowledgements}

Work reported here was supported in part by NSF grant 0748336.

\end{document}